\documentclass[aip,journal=cha,preprint,amsmath,amssymb,nofootinbib]{revtex4-2}
 
\usepackage{graphicx}
\usepackage{csquotes}
\usepackage[utf8]{inputenc}
\usepackage[T1]{fontenc}
\usepackage{multirow,bigdelim}
\newcommand{\vecg}[1]{\boldsymbol{#1}}
\newcommand{\tens}[1]{\mathbf{\underline{#1}}}
\newcommand{\tensg}[1]{\boldsymbol{\underline{#1}}}
\renewcommand\Re{\mathop{\textrm{Re}}}
\renewcommand\Im{\mathop{\textrm{Im}}}

\usepackage[usenames,dvipsnames]{color}
\graphicspath{{img/}}

\begin{document}

\title[An amplitude equation for the conserved-Hopf bifurcation]{An amplitude equation for the conserved-Hopf bifurcation - derivation, analysis and assessment}

\author{Daniel Greve}
\email{daniel.greve@uni-muenster.de}
\affiliation{Institut für Theoretische Physik, Universität Münster, Wilhelm-Klemm-Str. 9, 48149 Münster, Germany}
\author{Uwe Thiele}%
\email{u.thiele@uni-muenster.de}
\affiliation{Institut für Theoretische Physik, Universität Münster, Wilhelm-Klemm-Str. 9, 48149 Münster, Germany}
\affiliation{Center for Nonlinear Science (CeNoS), Universität Münster, Corrensstr. 2, 48149 Münster, Germany}
\affiliation{Center for Multiscale Theory and Computation (CMTC), Universität Münster, Corrensstr. 40, 48149 Münster,Germany}

\begin{abstract}
  We employ weakly nonlinear theory to derive an amplitude equation for the conserved-Hopf instability, i.e., a generic large-scale oscillatory instability for systems with two conservation laws. The resulting equation represents in the conserved case the equivalent of the complex Ginzburg-Landau equation obtained in the nonconserved case as amplitude equation for the standard Hopf bifurcation.

Considering first the case of a relatively simple symmetric Cahn-Hilliard model with purely nonreciprocal coupling, we derive the nonlinear nonlocal amplitude equation with real coefficients and show that its bifurcation diagram and time evolution well agree with results for the full model. The solutions of the amplitude equation and their stability are analytically obtained thereby showing that in oscillatory phase separation the suppression of coarsening is universal. Second, we lift the two restrictions and obtain the amplitude equation in the generic case that has complex coefficients, that also shows very good agreement with the full model as exemplified for some transient dynamics that converges to traveling wave states. 

\end{abstract}

\maketitle

\textbf{Nonreciprocal interactions and conservation laws both play an important role in out-of-equilibrium pattern formation processes, e.g., in biochemical systems. The generic oscillatory instability in such systems -- the conserved-Hopf instability -- is increasingly being studied as a central organizing element for the ongoing dynamic patterning. We employ a weakly nonlinear multi-scale analysis to obtain closed-form (but nonlocal) slow-time evolution equations for the spatio-temporal dynamics of the amplitude of fast-time oscillations for two-species Cahn-Hilliard models with nonreciprocal interactions. A comparison of analytical results for the reduced model and numerical results for the full model reveals excellent agreement. The derived amplitude equations will allow for a systematic study of universal behavior occurring in the vicinity of conserved-Hopf instabilities.}

\section{Introduction}\label{sec:intro}

For spatially extended dynamical systems, multiscale analyses represent a powerful tool to identify universal behavior close to the onset of linear instabilities, i.e., in the vicinity of local bifurcations.\cite{Craw1991rmp,CrHo1993rmp,Hoyle2006,Pismen2006} Independently of the specific system or instability, this relies on two key observations:  On the one hand, the growth rates of the mode(s) destabilized at the bifurcation are small, i.e., their dynamics occurs on a much slower timescale than the one of the stable modes. The latter relax on a fast timescale implying that they are slaved to the slow modes.\cite{Haken2012} On the other hand, in pattern forming systems unstable modes only occur in a small band of wavenumbers about the critical one. This results in the emergence of length scales and typically results in a scale separation between the length scale of a pattern and the length scale over which its amplitude or phase are modulated. \cite{CrHo1993rmp,Hoyle2006} 

The ensuing scale separation in space and/or time is then employed through various multiscale expansions, for systems that develop spatial structures or patterns resulting in amplitude (or envelope, or modulation) equations. These represent crucial simplifications as they are of reduced complexity with reduced order parameter and control parameter sets due to the eliminated small length and/or fast time scales. This can dramatically reduce the numerical effort. Often they even allow for analytical insights into universal behavior beyond linear approximations. Further, 
amplitude equations are quite valuable tools if one aims at a deeper understanding of possible generic qualitative behaviors in classes of systems and the changes occurring with changing parameters. This then allows one, for instance, to draw parallels between seemingly unrelated physical systems, here, e.g., between reaction-diffusion systems with two conservation laws and heated two-layer liquid films. However, also note that amplitude equations are a less than ideal tool if one aims at detailed quantitative predictions of nonlinear behavior for a particular, e.g. biophysical or material science, system at diverse sets of parameters far from instability onset.

A prominent classification of instabilities in spatially extended systems by Cross and Hohenberg \cite{CrHo1993rmp} distinguishes three types (I, II and III) with two respective flavors: stationary (or monotonic) and oscillatory -- in total, six categories of generic codimension-one instabilities. However, it has turned out that this traditional classification is not ideally suited when systematically considering systems with conservation laws that have come into sharper focus over the past two decades. \cite{MaCo2000n,WiMC2005n,IsOM2007pre,TARG2013pre,RAEB2013prl,BDGY2017nc,HaFr2018np,HaBF2018ptrsbs,BeGY2020c,YoFB2022prl,FrTh2023prl}

In consequence, a recent alternative classification into eight categories is based on three dichotomous properties, namely, large scale (L) vs.\ small scale (S) instability, stationary (s) vs.\ oscillatory (o) instability, and nonconserved (N) vs.\ conserved (C) dynamics of relevant mode(s). A table that lists the resulting eight instability types, the employed naming convention and their relation to the traditional classification is given in the supplementary material of Ref.~\onlinecite{FrTh2023prl}. Here, it is reproduced in appendix~\ref{app:table} as Table~\ref{tab:linstab} together with some further remarks.

Here, we solely focus on length and time scales that arise from the dispersion relations and their implications for the corresponding amplitude equation obtained by a weakly nonlinear approach. They are well known for the four cases without conservation law: Turing (small-scale, stationary) and Hopf (large-scale, oscillatory) instability give rise to the real and complex Ginzburg-Landau equation, respectively, that effectively describe the respective spatio-temporal modulation of harmonic spatial modulations and of harmonic temporal oscillations on large length- and slow timescale. Because no fast time or small length scale is induced by the dispersion relation, in the case of the Allen-Cahn instability (large-scale, stationary) the relevant \enquote{amplitude} is a simple scalar order parameter field, resulting in the Allen-Cahn equation as \enquote{amplitude equation}. In the case of a wave instability (small-scale, oscillatory) the amplitude equation corresponds to coupled complex Ginzburg-Landau equations for the amplitudes of left- and right-traveling waves. They are, however, only valid at small group velocity (see, e.g., section~VI.E of Ref.~\onlinecite{ArKr2002rmp}). 
Otherwise, the amplitude equation is nonlocal.\cite{Knob1992,Schn1997PRSEAM} The four basic cases without conservation law are all well covered by the Cross-Hohenberg classification and extensively analyzed in the literature.

However, this is less so for the four cases with conservation laws. There, one has to additionally account for the slow relaxation of the large-scale modes, i.e., with near-zero wavenumber. In the case of the conserved-Turing instability the real Ginzburg-Landau equation then couples to a nonlinear diffusion equation,\cite{MaCo2000n} while the case of the conserved-wave instability results in two complex Ginzburg-Landau equations coupled to an nonlinear diffusion equation. \cite{WiMC2005n} The Cahn-Hilliard instability is the conserved equivalent of the Allen-Cahn instability, implying that the relevant amplitude is also  a simple scalar field and the amplitude equation is a Cahn-Hilliard equation,\cite{BeRZ2018pre} explaining the coarsening behavior for many reaction-diffusion systems with one conservation law, in particular, in the vicinity of a large-scale stationary instability. \cite{IsOM2007pre,BeZi2019po,BrHF2020prx,BWHY2021prl}

In the final case of a conserved-Hopf instability, two conservation laws are needed to allow for a large-scale oscillatory instability of a conserved mode. Although such instabilities and the resulting nonlinear oscillatory behavior have been described for a number of systems with at least two conservation laws \cite{Wola2002ejam,JoBa2005pb,PoTS2016epje,RAEB2013prl,NeSh2016jpat,NeSi2017pf,HaFr2018np,TYDF2022sa} a systematic understanding of the underlying universal behavior has still to emerge. A particular case that has recently attracted considerable attention is the nonreciprocally coupled Cahn-Hilliard model. \cite{YoBM2020pnasusa,SaAG2020prx,FrWT2021pre,FrTh2021ijam,FrTP2023ptrsapes,FHKG2023pre,FrTh2023prl,GLFT2024preprint,SuKL2023prl,SuKL2023pre,SuKL2023preb,AlCB2023prl,BrMa2024prx}

Ref.~\onlinecite{FrTh2023prl} has shown that a general nonreciprocal two-component Cahn-Hilliard model captures the universal behavior in the vicinity of any conserved-Hopf instability in terms of the dynamics of two density-like scalar fields that directly correspond to the trivial amplitudes of the zero-modes as in the corresponding stationary case. However, although the resulting \enquote{amplitude equation} captures the corresponding behavior of many specific systems and encodes their parameters in its universal but still extremely rich parameter set it does not correspond to an amplitude equation in the spirit of a Ginzburg-Landau-type equation that eliminates a small length or fast time scale. Actually, the equation derived in Ref.~\onlinecite{FrTh2023prl} instead corresponds to the amplitude equation for an instability of higher codimension because it encompasses further primary bifurcations beside the conserved-Hopf instability, e.g., a codimension-two Cahn-Hilliard instability. Therefore, the corresponding proper amplitude equation that eliminates the time scale of fast oscillations is still missing.

Our present work develops such an amplitude equation and in this way hopefully completes the set of eight such equations for the eight above discussed linear instabilities. Our approach is based on a close inspection of the dispersion relation that implies the occurrence of two different timescales, see Fig.~\ref{fig:disp_cons_hopf}. On the one hand, there are the frequencies of the unstable modes. They scale with $\varepsilon^2$ where $\varepsilon$ represents the distance to the onset of the conserved-Hopf instability. On the other hand, the fastest growing unstable mode has a growth rate of order $\varepsilon^4$. We show that these timescales can be separated, ultimately resulting in an amplitude equation that captures the nonlinear behavior of the slowly evolving amplitude of the modes with $\varepsilon^2$-frequencies in the vicinity of the conserved-Hopf bifurcation.

This work is structured as follows. In section~\ref{sec:basics} we discuss the scaling properties resulting from the dispersion relation of the conserved-Hopf bifurcation, and introduce the relatively simple but instructive example of a symmetric Cahn-Hilliard model with purely nonreciprocal linear coupling. We identify the resulting linear problem as a free-particle Schr\"odinger equation and discuss the resulting multi-scale ansatz in real and Fourier space. Subsequently, in section~\ref{sec:ae}, we derive the leading-order amplitude equation and briefly discuss its structure and properties. In section~\ref{sec:ae_sol_analytic} we proceed by determining analytic solutions of the obtained amplitude equation, their linear stability and resulting bifurcation structure. The discussion carries over to 
section~\ref{sec:ae_sol_nuemeric} where we quantitatively compare states, bifurcation structure and time evolution obtained from the amplitude equation and from the symmetric purely nonreciprocally coupled fully nonlinear model. Finally, we lift the restrictions imposed by the simplicity of the case considered in sections~\ref{sec:basics} to~\ref{sec:ae_sol_nuemeric} and extend the developed approach towards the generic case.
We summarize and conclude in section~\ref{sec:conclusion}.

\section{Scaling, ansatz and symmetry}\label{sec:basics}

\begin{figure}[htb!]
\centering
	\includegraphics[width=0.7\hsize]{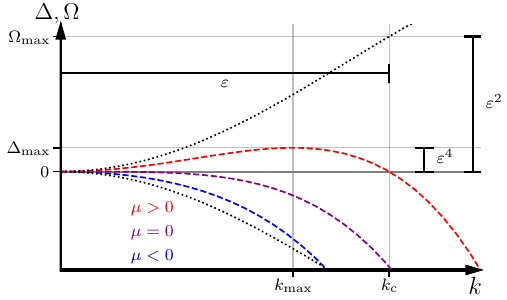}
	\caption{Dispersion relation of the conserved Hopf instability. The growth rate $\Delta$ is respectively shown in blue, purple and red below ($\mu<0$), at ($\mu=0$) and above ($\mu>0$) the onset of instability. The dotted lines indicate the corresponding frequencies $\pm \Omega$.} 
	\label{fig:disp_cons_hopf}
\end{figure}

We consider a system of spatially and temporally homogeneous and parity symmetric kinetic equations of first order in time for scalar fields. For simplicity, we consider a large, one-dimensional domain of size $l$ with periodic boundary conditions.  The Taylor-expanded dispersion relation for a conserved-Hopf bifurcation is \cite{FrTh2023prl}
\begin{equation}
\begin{split}\label{eq:dispersion}
	\lambda_{\pm}(k)&=\Delta(k)\pm i\Omega(k)\\
	\Delta(k)&=\mu k^2 -\delta' k^4+\mathcal{O}(k^6)\\
	\Omega(k)&=\omega k^2 - \omega'k^4+\mathcal{O}(k^6),\\
\end{split}	
\end{equation}
where $\mu=\varepsilon^2 \tilde{\mu} \ll 1$,  $\tilde{\mu}=\mathcal{O}(1)$ and $\omega=\mathcal{O}(1)$. In other words, the system is close to the onset of a conserved, large-scale, oscillatory instability, where for $\tilde{\mu}>0$ modes with complex conjugate pairs of eigenvalues with $k<k_{\mathrm{c}}=\varepsilon\sqrt{\tilde{\mu}/\delta'}$ will exponentially grow whereas all other modes will decay exponentially, see Fig.~\ref{fig:disp_cons_hopf}. 
A simple example, that, similar to the Swift-Hohenberg model for the Turing instability, directly and without Taylor-expansion yields the dispersion relation~\eqref{eq:dispersion}, is the symmetric Cahn-Hilliard model with purely nonreciprocal linear coupling 
\begin{equation}\label{eq:minimal_CH}
\begin{split}
	\partial_t u&=\partial_{xx}(-\mu u+u^3-\partial_{xx}u- \omega v)\\
	\partial_t v&=\partial_{xx}(-\mu v+v^3-\partial_{xx}v+\omega u),\\[.5em]
\text{in the particular case}&\quad \int_0^l u \mathrm{d}x =\int_0^l v \mathrm{d}x =0. 
\end{split}
\end{equation}
This is a special case of the models analyzed in Refs.~\onlinecite{FrTP2023ptrsapes,FHKG2023pre,FrWT2021pre,FrTh2021ijam,SaAG2020prx,YoBM2020pnasusa,AlCB2023prl}, where sometimes noise is added. Further simplified versions are obtained by linearizing the second equation. \cite{Foertsch2023Bayreuth,SuKL2023pre,SuKL2023preb,BrMa2024prx} Using a vector notation $\vecg u=(u,v)$ and employing the usual ansatz $\vecg u(t)=\vecg u_0+\delta \vecg u(t)$ with $\delta \vecg u(t)\sim \, \varepsilon e^{\mathrm{i}kx-\lambda t}\ll 1$ in \eqref{eq:minimal_CH}, i.e., adding a small perturbation $\delta \vecg u$ to the trivial solution $u(x,t)=v(x,t)=0$, we obtain the dispersion relation \eqref{eq:dispersion} with $\delta'=1$ and $\omega'=0$. 

Next, we observe that the fastest growing wavenumber $k_{\text{max}}$ is of $\mathcal{O}(\varepsilon)$ which implies that the dynamics occurs on the length scale $X=\varepsilon x$. This implies that the leading order oscillation in \eqref{eq:dispersion} occurs on the time scale\footnote{Note, that one could similarly argue for the frequency $\Omega(k_{\mathrm{max}})$ at the largest growing wavenumber $k_{\mathrm{max}}$ to be used for the fast timescale. However, since $k_{\mathrm{max}}=k_c/\sqrt{2}$, this  does not affect the scaling in $\varepsilon$.} $\tau=\varepsilon^2 t$, whereas the first contribution of the growth rates occur on the timescale $T=\varepsilon^4 t$. Then, we perform a weakly nonlinear analysis for the symmetric model \eqref{eq:minimal_CH} with the scalings $\partial_x=\varepsilon \partial_{X}$ and $\partial_t=\varepsilon^2 \partial_\tau +\varepsilon^4 \partial_T$ and

\begin{equation}\label{eq:field_expansion}
\vecg u(t)=\varepsilon \vecg u_1(X,\tau,T)+\varepsilon^2 \vecg u_2(X,\tau,T)+\varepsilon^3 \vecg u_3(X,\tau,T)+\text{h.o.t.}.
\end{equation}
The first nonvanishing contribution occurs at $\mathcal{O}(\varepsilon^3)$ and yields the homogeneous linear problem
\begin{equation}\label{eq:Schrödinger_eq}
\begin{split}
	\tens L \vecg u_1 &= 0 \\
\text{with}\quad	\tensg{L}&= \tensg{1}\partial_\tau-\begin{pmatrix}
	 0&-\omega\\
	\omega & 0
	\end{pmatrix}\partial_{XX}\\
\end{split}
\end{equation}
This corresponds, up to prefactors, to the linear, i.e., free-particle Schrödinger equation\footnote{The fields $\vecg u$ correspond to the real and imaginary part of a wave function $\psi$, i.e., the free-particle Schrödinger equation in its common form can be brought into this form via adding and subtracting its complex conjugate:
\begin{equation*}
	i\partial_t \psi =-\partial_{xx}\psi \qquad \Leftrightarrow\qquad
	\partial_t \begin{pmatrix}
	\Im \psi\\\Re \psi
	\end{pmatrix}=\begin{pmatrix}
	0&1\\-1&0
	\end{pmatrix}\partial_{xx}\begin{pmatrix}
	\Im \psi\\\Re \psi
	\end{pmatrix}
\end{equation*}
}. Its general solution is
\begin{equation}\label{eq:ansatz}
\vecg u_1(X,\tau,T)=e^{i\omega \tau \partial_{XX}} A(X,T) \vecg e_+ + e^{-i\omega \tau \partial_{XX}} A^*(X,T) \vecg e_-, 
\end{equation}
where $\vecg e_{\pm}=(\pm i,1)/\sqrt{2}$ (i.e., $\vecg e_+\perp \vecg e_-$). In other words, $e^{\mathrm{i}\omega\tau \partial_{XX}}$ corresponds to a time evolution operator, where $-\omega \partial_{XX}$ is the free-particle Hamiltonian and the slowly varying amplitude $A(X,T)$ takes to role of the initial wavefunction that (on the fast timescale $\tau$) freely propagates in space. Note, that the ansatz breaks down if $\omega=0$ or $\omega=\mathcal{O}(\varepsilon^2)$, as then Eq.~\eqref{eq:Schrödinger_eq} becomes trivial ($\partial_\tau \vecg u_1 =0$) and all terms enter the expansion at $\mathcal{O}(\varepsilon^5)$, i.e., no separation of time scales is possible. This occurs, e.g., when studying the codimension-two point where a transition between Cahn-Hilliard instability and conserved Hopf instability occurs. There, $\omega\rightarrow 0$ and the critical wave number is zero for both instabilities. Therefore, considering such a transition, the nonreciprocal Cahn-Hilliard (NRCH) model cannot be reduced by scale separation and therefore emerges as its own amplitude equation, see Ref.~\onlinecite{FrTh2023prl}.

Next, we discuss the consequences of the ansatz \eqref{eq:ansatz} that arises from the linear problem at leading order. In general, the underlying symmetry of the original system, i.e., parity symmetry and homogeneity in space and time combined with the relation between amplitude and original variables implies consequences for the form of occurring nonlinearities. For instance, for the standard Hopf instability without conservation law, the usual ansatz is $\vecg u(x,t)=\varepsilon e^{\mathrm{i} \omega t} A(X,\tau)\vecg e_++c.c.$. Then, homogeneity in time implies that the equations are invariant under the shift $t\rightarrow t+t_0$. Hence, on the amplitude level, the amplitude $A$ picks up a global phase $A(X,\tau)\rightarrow e^{\mathrm{i}\omega t_0}A(X,\tau)$. From this we can conclude that the simplest occurring nonlinearity is $\sim A|A|^2$ as it transforms like the linear term, i.e., $A|A|^2\rightarrow e^{\mathrm{i}\omega t_0}A|A|^2$, whereas quadratic or other cubic terms are forbidden, because e.g., $A^2\rightarrow e^{2i\omega t_0}A^2$ shows a different transformation behavior.\footnote{Analogously, for Turing instabilities, where $\vecg u(x,t)=\varepsilon e^{\mathrm{i} k x} A(X,\tau)\vecg e_++c.c..$, the combination of homogeneity in space with additional parity invariance yields that the simplest nonlinearity is $\sim A|A|^2$ with a real coefficient.}

We transfer this argument to the ansatz \eqref{eq:ansatz} for the conserved Hopf instability. Here and in the following, it is instructive to discuss the problem in terms of Fourier modes, i.e., we introduce the usual Fourier transform $\tilde{A}(K,T)$, which fulfils $A(X,T)=\sum_K \tilde{A}(K,T) e^{ \mathrm i KX}$, where $K=2 \pi z /L$, with $L=\varepsilon l$ and $z\in \mathbb{Z}$. Hence, the amplitude transforms under $\tau\rightarrow \tau+\tau_0$ as
\begin{equation}\label{eq:amplitude}
\begin{split}
	A(X,T)&\rightarrow e^{\mathrm{i}\omega\tau_0 \partial_{XX}}A(X,T)\\
	&=e^{\mathrm{i}\omega\tau_0 \partial_{XX}} \sum_K \tilde{A}(K,T) e^{ \mathrm i KX}\\
	&=\sum_K e^{-\mathrm{i}\omega\tau_0 K^2} \tilde{A}(K,T) e^{  \mathrm i KX}.
	\end{split}
\end{equation}
In other words, by the convolution theorem each Fourier mode $\tilde{A}(K,T)$ of the amplitude transforms individually as $\tilde{A}(K,T)\rightarrow e^{-\mathrm{i}\omega\tau_0 K^2}\tilde{A}(K,T)$. This has dramatic consequences for nonlinearities of the amplitude equation. Especially, it rules out all local nonlinearities, as they generically couple distinct Fourier modes with nonuniform transformation behavior. However,  (nonlocal) nonlinearities compatible with the symmetry can be explicitly constructed based on their Fourier space representation, like, e.g., $	\mathcal{F}^{-1}[\tilde{A}|\tilde{A}|^2]$. 
Next, we will derive the amplitude equation, i.e., a dynamical equation for $A(X,T)$ at leading order.

\section{Leading order amplitude equation: simple case}\label{sec:ae}
At $\mathcal{O}(\varepsilon^4)$ we obtain a similar homogeneous linear system for $\vecg{u}_2$ as Eqs.~\eqref{eq:Schrödinger_eq} for $\vecg{u}_1$ at $\mathcal{O}(\varepsilon^3)$. Here, it is not considered further as it does not contribute at $\mathcal{O}(\varepsilon^5)$. At $\mathcal{O}(\varepsilon^5)$, we obtain the inhomogeneous linear system
\begin{equation}\label{eq:O_epsilon_5}
\begin{split}
	\tensg {L} \vecg u_3&=\vecg q_5 \quad\text{with} \\
	 \vecg q_5&=\partial_{XX}\begin{pmatrix}
	-\tilde{\mu} u_1+u_1^3-\partial_{XX}u_1\\
	-\tilde{\mu} v_1+v_1^3-\partial_{XX}v_1\\
	\end{pmatrix}-\begin{pmatrix}\partial_T u_1\\\partial_T v_1\end{pmatrix}. 
\end{split}
\end{equation}
To apply a Fredholm alternative, we define a \enquote{spatiotemporal scalar product} in the function space of spatially $L$-periodic and temporally $1/\Omega$-periodic functions
\begin{equation}\label{eq:scalar_prod}
	\langle \vecg F;\vecg G \rangle = \frac{\Omega_{\mathrm{min}}}{L}\int\limits_{0}^{L} \mathrm{d} X \int\limits_0^{1/\Omega_{\mathrm{min}}} \mathrm{d} \tau \vecg F^\dagger\vecg G,
\end{equation}
where $\Omega_{\min} =\omega K_{\mathrm{min}}^2/(2\pi)=2\pi \omega /L^2$. Then, the integral of a harmonic function is only nonzero, i.e., $\frac{\Omega_{\mathrm{min}}}{L}\int_0^L \mathrm{d} X \int_0^{1/\Omega_{\mathrm{min}}} \mathrm{d} \tau e^{\mathrm{i}KX} e^{\mathrm{i}\Omega\tau}=1$ if both spatial and temporal frequencies vanish $(K=\Omega=0)$. It is zero otherwise.

By the Fredholm alternative, Eq.~\eqref{eq:O_epsilon_5} only has a solution if $\vecg q_5$ is orthogonal to the kernel of the adjoint linear operator $\tensg L^\dagger$, given by
\begin{equation}
	\tensg {L}^\dagger=-\tensg{1}\partial_\tau -\begin{pmatrix}
	 0&\omega\\
	-\omega& 0
	\end{pmatrix}\partial_{XX}=-\tensg L. 
\end{equation}
In line with Eqs.~\eqref{eq:Schrödinger_eq} and \eqref{eq:ansatz} at $\mathcal{O}(\varepsilon^3)$ the kernel of this linear operator is given by
\begin{equation}\label{eq:fredh_kernel}
	\vecg m = e^{\mathrm{i}\omega\tau\partial_{XX}}B(X)\vecg e_++e^{-\mathrm{i}\omega\tau \partial_{XX}}C(X)\vecg e_-,
\end{equation}

where $B(X)$ and $C(X)$ are arbitrary functions. To obtain an equation for the amplitude $A(\hat{X},T)$ we choose\footnote{In principle other choices are possible. While exchanging the roles of $B(X)$ and $C(X)$ results in the equation for $A^*(X)$, keeping $B(X)$ arbitrary and $C(X)=0$ will result in an evolution equation of type $\partial_T \langle	B(X),A(X,T) \rangle = \dots	$.  One could instead choose harmonics, i.e., $B(X)\sim e^{\mathrm{i}QX}$ to first calculate projections of the amplitude onto single Fourier modes, and then construct the amplitude as their superposition. However, our choice already makes use of the complete null space, since the set of functions used for $B(X)$, i.e., $\{\delta(X-\hat{X});\hat{X}\in [0,L]\}$ represents already a complete basis of the function space of periodic functions in $[0,L]$ and we hence already made use of the orthogonality to every element in the (infinite dimensional) null space of $\tensg L^\dagger$.} $B(X)=\delta(X-\hat{X})$ and $C(X)=0$. \newpage

Then, applying the Fredholm alternative for the linear contributions in $\vecg q_5$, i.e., evaluating their scalar product \eqref{eq:scalar_prod} with Eq.~\eqref{eq:fredh_kernel}  and substituting $\vecg u_1$ with the amplitude \eqref{eq:ansatz}, we obtain
\begin{equation}
\begin{split}
&\langle \vecg m;(-\tilde{\mu}\partial_{XX}- \partial_{XXXX}-\partial_T) \vecg u_1 \rangle\\
 =&\langle e^{\mathrm{i}\omega\tau\partial_{XX}}\delta(X-\hat{X})\vecg e_+;(-\tilde{\mu}\partial_{XX}- \partial_{XXXX}-\partial_T)\\
 &(e^{\mathrm{i}\omega\tau\partial_{XX}}A(X,T)\vecg e_+ +e^{-\mathrm{i}\omega\tau\partial_{XX}}A^*(X,T)\vecg e_-) \rangle\\
=&(-\tilde{\mu}\partial_{\hat{X}\hat{X}}- \partial_{\hat{X}\hat{X}\hat{X}\hat{X}}-\partial_T)A(\hat{X},T),
\end{split}
\end{equation}
where we employed the orthogonality of $\vecg e_+$ and $\vecg e_-$, and used the unitarity\footnote{
Note that exponentiating $\mathrm{i}H$  always results in the unitarity of $e^{\mathrm{i}H}$ for a hermitian operator $H$. Further, here we can easily proof unitarity using Parseval's theorem, i.e.,\begin{equation*}
\begin{split}
	\langle e^{\mathrm{i}\omega \tau\partial_{XX}} A(X), e^{\mathrm{i}\omega \tau \partial_{XX}} B(X) \rangle
	&=\frac{1}{L}\int_0^L\mathrm{d}X \left[e^{-\mathrm{i}\omega \tau \partial_{XX}} A^\dagger(X)\right]\left[e^{\mathrm{i}\omega \tau \partial_{XX}} B(X)\right]\\
&=\sum_K \left[e^{\mathrm{i}\omega \tau K^2}\tilde{A}^\dagger(K)\right]\left[e^{-\mathrm{i}\omega \tau K^2} \tilde{B}(K)\right]=\sum_K \tilde{A}^\dagger(K) \tilde{B}(K)\\
&=\frac{1}{L}\int_0^L\mathrm{d}X A^\dagger(X)B(X)=\langle A(X), B(X) \rangle.
\end{split}
\end{equation*}} of the operator $e^{-\mathrm{i}\omega\tau\partial_{XX}}$. Next, we consider the cubic nonlinearities within $\vecg q_5$. Inserting the amplitude $A(X,T)$ from \eqref{eq:ansatz} yields
\begin{equation}\label{eq:ampl_nonl_u}
\begin{split}
\partial_{XX}&\begin{pmatrix}u_1^3\\v_1^3\end{pmatrix}
= \partial_{XX}\biggl[\left(e^{\mathrm{i}\omega\tau\partial_{XX}} A(X,T)\right)^3\vecg w^{(1)}\\
+&3\left(e^{-\mathrm{i}\omega\tau\partial_{XX}} A^*(X,T)\right)\left(e^{\mathrm{i}\omega\tau\partial_{XX}} A(X,T)\right)^2 \vecg w^{(2)}\\
+&3\left(e^{-\mathrm{i}\omega\tau\partial_{XX}} A^*(X,T)\right)^2\left(e^{\mathrm{i}\omega\tau\partial_{XX}} A(X,T)\right)\vecg w^{(3)}\\
+&\left(e^{-\mathrm{i}\omega\tau\partial_{XX}} A^*(X,T)\right)^3\vecg w^{(4)}\bigg].
\end{split}
\end{equation}
The $\vecg w^{(i)}$ are given by
\begin{equation}
\begin{split}
\vecg w^{(1)}&=\begin{pmatrix}
(\vecg e_+)_1^3 \\ 
(\vecg e_+)_2^3 
\end{pmatrix},\quad 
\vecg w^{(2)}=\begin{pmatrix}
(\vecg e_+)_1^2 (\vecg e_-)_1\\ 
(\vecg e_+)_2^2 (\vecg e_-)_2
\end{pmatrix}\\ 
\vecg w^{(3)}&=\begin{pmatrix}
(\vecg e_+)_1 (\vecg e_-)_1^2\\ 
(\vecg e_+)_2 (\vecg e_-)_2^2
\end{pmatrix} \quad \text{and}\quad 
\vecg w^{(4)}=
 \begin{pmatrix}
(\vecg e_-)_1^3 \\ 
(\vecg e_-)_2^3 
\end{pmatrix}. 
\end{split}
\end{equation}
where $(\vecg e_\pm)_1$ and $(\vecg e_\pm)_2$ refer to the first and second component of $\vecg e_\pm$, respectively. For the presently considered simple nongeneric choice of linear and nonlinear terms, we obtain $\vecg w^{(1)}=\vecg w^{(3)}=\vecg e_-/2$ and $\vecg w^{(2)}=\vecg w^{(4)}=\vecg e_+/2$. 

First, we deal with the second term $\sim \vecg w^{(2)}$ in \eqref{eq:ampl_nonl_u}, as it results in the dominant nonlinear contribution.  We use the identity $\delta(X-\hat{X})=\sum_Q e^{\mathrm{i}Q(X-\hat{X})}$ and again express all occurring amplitudes by their Fourier transform $A(X,T)=\sum_K \tilde{A}(K,T)e^{\mathrm{i}KX}$ to evaluate the scalar product
{\small \begin{widetext} 
\begin{equation}
\begin{split}\label{eq:nonlinearity_proj}
&3\left\langle e^{\mathrm{i}\omega\tau\partial_{XX}}\delta(X-\hat{X})\vecg e_+;\partial_{XX}\left(e^{-\mathrm{i}\omega\tau\partial_{XX}} A^*(X,T)\right)\left(e^{\mathrm{i}\omega\tau\partial_{XX}} A(X,T)\right)^2 \vecg w^{(2)}\right\rangle\\
=&\frac{3}{2}\left\langle e^{\mathrm{i}\omega\tau\partial_{XX}}\sum_Q e^{\mathrm{i}Q(X-\hat{X})}\vecg e_+;\partial_{XX}\left(e^{-\mathrm{i}\omega\tau\partial_{XX}} \sum_K \Tilde{A}^*(K,T)e^{\mathrm{i}KX}\right)\left(e^{-\mathrm{i}\omega\tau\partial_{XX}} \sum_K \Tilde{A}(K,T)e^{\mathrm{i}KX}\right)^2 \vecg e_+\right\rangle\\
=&\frac{3 \Omega_{\text{min}}}{2L}\sum_{Q,K,K',K''} e^{\mathrm{i}Q\hat{X}} \Tilde{A}^*(K,T)\Tilde{A}(K',T)\Tilde{A}(K'',T) \times\\ & \times\left(-Q^2\right)\int_0^{1/\Omega_{\text{min}}} e^{\mathrm{i}\omega\tau (Q^2+K^2-K'^2-K''^2)} \mathrm{d} \tau \int_0^L e^{\mathrm{i}X(-Q-K+K'+K'')} \mathrm{d} X \\
	=&\frac{3}{2}\left[2\sum_Q e^{\mathrm{i}Q\hat{X}} (-Q^2)\tilde{A}(Q,T) \sum_K \tilde{A}^*(K,T) \tilde{A}(K,T)-\sum_{Q} e^{\mathrm{i}Q\hat{X}} (-Q^2)\tilde{A}(Q,T) \tilde{A}(Q,T) \tilde{A}^*(Q,T)\right]\\
	=&\frac{3}{2}\partial_{\hat{X}\hat{X}}\left(2  \mathcal{F}^{-1}[\tilde{A}(K,T)\sum_{K'}|\tilde A(K',T)|^2]-\mathcal{F}^{-1}[\tilde{A}(K,T)|\tilde A(K,T)|^2] \right).
\end{split}
\end{equation}
\end{widetext}
}
Here, the key observation is that the two integrals in the third line are zero, whenever one of the integrands has a nonzero phase. Therefore only terms with
\begin{equation}\label{eq:solv_cond}
Q^2+K^2=K'^2+K''^2 \quad \text{and} \quad Q+K=K'+K''
\end{equation}
contribute.  Eqs.~\eqref{eq:solv_cond} have a strong analogy to an elastic collision, i.e., they correspond to the nonrelativistic, elastic $2\rightarrow 2$ scattering of particles with equal mass, incoming momenta $Q,K$  and outgoing momenta $K',K''$. In one dimension, the only possibilities are that the two particles exchange momenta or pass unhindered, i.e., the momenta are pairwise equal $Q=K'$ and $K=K''$ or $Q=K''$ and  $K=K'$. After inserting into Eq.~\eqref{eq:nonlinearity_proj}, both options yield the same contribution. However, if $Q=K$ the two solutions are identical, i.e., we have to account for overcounting, which is done by the second term in the fourth row of Eq.~\eqref{eq:nonlinearity_proj}.
Due to Parseval's identity, the first term transforms into real space, as $\mathcal{F}^{-1}[\tilde{A}(K,T)\sum_{K'}|\tilde A(K',T)|^2]=A(X,T)\langle|A(X,T)|^2\rangle$, i.e., as a coupling to the mean of the squared amplitude. In real space the second term can be expressed as the double-convolution $\mathcal{F}^{-1}[\tilde{A}|\tilde A|^2](X)=[A(\tilde{X})\star A(\tilde{X})\star A^*(-\tilde{X})](X)$, but we will keep the first notation for brevity. 
A similar consideration for the three remaining nonlinearities yields that they correspond to elastic $4\rightarrow 0$,  $3\rightarrow 1$ and $1\rightarrow 3$ scatterings of particles with equal mass\footnote{Here, elastic means that energy and momentum are both conserved. However, mass is not conserved, e.g., for the $1\rightarrow 3$ case, one particle splits into three particles, where each has identical mass as the incident particle.} in one dimension. Trivial solutions, where some of the momenta equal zero do not contribute, since $A(K=0,T)=0$ which results from the mass conservation of the underlying system. Further, the latter two contributions are also negligible compared to the $2 \rightarrow 2$ scattering if the amplitude is sufficiently localized in Fourier space, which we assume. 
Finally, collecting all terms and reintroducing the original scales, i.e., introducing $a=\varepsilon A$, we obtain the amplitude equation

\begin{equation}\label{eq:AE_og_scales}
  \partial_t a=\partial_{xx}\left(-\mu a+\frac{3}{2}(2a\langle |a|^2\rangle-\mathcal{F}^{-1}[\tilde{a}|\tilde a|^2])-\partial_{xx}a\right),
\end{equation}
were the coefficient(s) are all real. The nature of the  nonlocal terms can be understood in terms of the dynamics on the different time scales: Due to the dispersion relation $\Omega(k)\sim \pm \omega k^2$ each individual harmonic mode propagates with a distinct velocity $\pm\omega k$ on the fast time scale $\tau$. As the propagation of the wave is very fast as compared to the growth of their amplitudes, during a characteristic time interval of amplitude growth, several wavelength of other waves will have passed, resulting in a nonlocal interaction. In other words each wave interacts with the (squared) mean of all other waves. This holds for all interactions besides the self-interaction of a wave, since in this case the propagation velocity trivially coincides.
The amplitude equation is related to the dynamics of the original fields $\vecg u$ via
\begin{equation}\label{eq:ae_relation_og_scales}
\quad \vecg{u}(x,t)=e^{\mathrm{i}\omega t\partial_{xx}}a(x,t)\vecg e_++e^{-\mathrm{i}\omega t\partial_{xx}}a^*(x,t)\vecg e_-
\end{equation}
The dispersion relation of equation \eqref{eq:AE_og_scales} is $\lambda(k^2)=\mu k^2-k^4$, i.e., compared to the general Eq.~\eqref{eq:dispersion} the frequencies $~\pm \omega k^2$ on the fast timescale do no longer appear, i.e., the fast timescale has been fully separated. In the considered case, the instability is even stationary, i.e., the eigenvalues $\lambda(k^2)$ are all real. However, this is a nongeneric consequence of our minimal choice \eqref{eq:minimal_CH}, i.e., without a contribution $\sim k^4$ to the frequency, i.e., $\omega'=0$ in Eq.~\eqref{eq:dispersion}. The general case is treated below in section~\ref{sec:ae_generic}.

Remarkably, the obtained amplitude equation with real coefficients has the structure of a gradient dynamics for a conserved complex field, i.e., it can be written in the form
\begin{equation}\label{eq:ae-graddyn}
	\partial_t a=\partial_{xx}\frac{\delta \mathcal{G}[a]}{\delta a^*},
\end{equation}
where the underlying energy functional is
\begin{equation}\label{eq:free_energy_ae}
\begin{split}
\mathcal{G}[a]&=\int\limits_{-l/2}^{l/2}\left(-\mu |a|^2+|\partial_x a|^2\right)\mathrm{d} x+\frac{3l}{2}\left(\frac{1}{l}\int\limits_{-l/2}^{l/2}|a|^2 \mathrm{d} x \right)^2
\\&-\frac{3}{4l^2}\iiint\limits_{[-\frac{l}{2},\frac{l}{2}]^3}a^*(x)a^*(-x')a(x'')a(x-x'-x'')\mathrm{d} x\,\mathrm{d} x'\mathrm{d} x''\,.
\end{split}
\end{equation}
The energy is more intuitively understood, if it is expressed by the Fourier amplitude $\tilde{a}$, where $a(x,t)=\sum_k \tilde{a}(k,t)e^{ikx}$. It reads
\begin{equation}\label{eq:energy-fourier}
	\mathcal{G}[\tilde{a}]=l\sum_k\left[ (-\mu+k^2) |\tilde{a}|^2-\frac{3}{4}|\tilde{a}|^4\right]+\frac{3l}{2}\left(\sum_k |\tilde{a}|^2\right)^2.
\end{equation}
As $(\sum_k |\tilde{a}|^2)^2\geq \sum_k |\tilde{a}|^4$, for large values of $\tilde{a}$ the energy is dominated by the strictly positive last term. This shows that it is bounded from below. The variational character of \eqref{eq:ae-graddyn} has important implications: The evaluation of the total time derivative of $\mathcal{G}[a]$, i.e.,
\begin{equation}
\begin{split}
\frac{\mathrm{d}}{\mathrm{d} t}\mathcal{G}[a]=&\int\limits_{-l/2}^{l/2}\left(\frac{\delta \mathcal{G}}{\delta a^*}\frac{\mathrm{d} a^*}{\mathrm{d} t} + \frac{\delta \mathcal{G}}{\delta a}\frac{\mathrm{d} a}{\mathrm{d} t}\right)\mathrm{d} x\\
=&\int\limits_{-l/2}^{l/2}\left[\frac{\delta \mathcal{G}}{\delta a^*}\left(\partial_{xx} \frac{\delta \mathcal{G}}{\delta a}\right)+\frac{\delta \mathcal{G}}{\delta a}\left(\partial_{xx} \frac{\delta \mathcal{G}}{\delta a^*}\right)\right]\mathrm{d} x\overset{\text{p.I.}}{=}-2\int\limits_{-l/2}^{l/2}\left|\partial_{x}\frac{\delta \mathcal{G}}{\delta a}\right|^2\mathrm{d} x\leq 0 \, ,
\end{split}
\end{equation}
demonstrates that $\mathcal{G}[a]$ is a Lyapunov functional. This implies that the amplitude $a(x,t)$ will always eventually approach a stable steady state, and all stable steady states will only show eigenmodes with real eigenvalues. This is, no oscillatory dynamics occurs on the level of the amplitude equation, although it does on the level of the original fast dynamics. 

Next, we will analyze the analytical solutions of the amplitude equation \eqref{eq:AE_og_scales} and relate them to the dynamics of the full model.
\section{States and their stability}\label{sec:ae_sol_analytic}
Before entering fine details, we discuss the simple harmonic solution to the amplitude equation~(\ref{eq:AE_og_scales}). As shown below it is given by
\begin{equation}\label{eq:harmonic_soluitons}
	a(x,t)=a_0e^{\mathrm{i}qx} \quad \text{with} \quad |a_0|^2=\frac{2}{3}(\mu-q^2),
\end{equation}
where $|q|\in(0,\sqrt{\mu})$ for $\mu>0$. Using Eq.~\eqref{eq:ae_relation_og_scales} to translate this back into the original fields, i.e., applying the linear Schrödinger-like time evolution yields
\begin{equation}
	\vecg u =\sqrt{2}|a_0|\begin{pmatrix}
	\sin(qx+\omega q^2 t+\phi)\\
	\cos(qx+\omega q^2 t+\phi)
	\end{pmatrix}.
\end{equation}
 Therefore a single-mode harmonic solution of the amplitude equation corresponds to a traveling wave with velocity $-\omega q$, where the fields $u_1$ and $u_2$ are phase shifted by $\pi/2$. Left-traveling [right-traveling] waves are hence given by $q>0$ [$q<0$]. The global phase $\phi$ is undetermined and reflects translational symmetry.  
 
\begin{figure}[htb!]
\centering
\includegraphics[width=0.7\hsize]{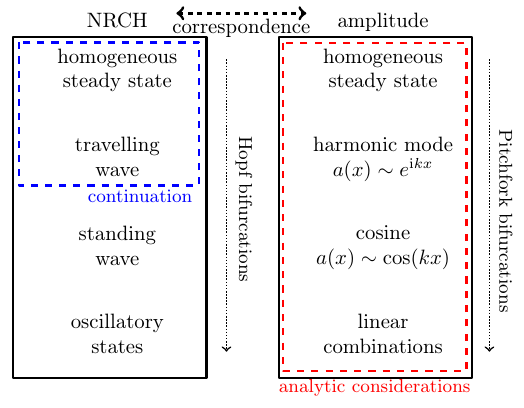}
\caption{Scheme of the corresponding states in the nonreciprocal Cahn-Hilliard  (NRCH) model \eqref{eq:minimal_CH} and the obtained amplitude equation with real coefficients \eqref{eq:AE_og_scales}. The blue rectangle highlights the states captured with numerical path continuation, whereas the right rectangle displays states that are obtained from analytic considerations in section~\ref{sec:ae_sol_analytic}. }
\label{fig:correspondence}
\end{figure} 

To find the general solution of the steady state equation, i.e., Eq.~(\ref{eq:AE_og_scales}) with $\partial_t a=0$, we integrate twice. The first integration constant is zero as we request the net flux to be zero. Remarkably, here we can also eliminate the second integration constant as we have to request $\overline{a}=0$, because the amplitude $a$ defines the perturbation with respect to the mean value of the original field. As a result we have
\begin{equation}
	-\mu a+3a \langle |a|^2\rangle-\frac{3}{2}\mathcal{F}^{-1}[\tilde{a}|\tilde{a}|^2]-\partial_{xx}a=0\, .
\end{equation}
To solve it, we reintroduce the Fourier-transform, i.e., we substitute $a(x,t)=\sum_q \tilde a(q)e^{\mathrm{\mathrm{i}}qx}$ to obtain
\begin{equation}
	\sum_q \left(-\mu +3\sum_k |\tilde{a}(k)|^2-\frac{3}{2}|\tilde{a}(q)|^2+q^2\right) \Tilde{a}(q)e^{\mathrm{i}qx}=0.
\end{equation}
Since the $e^{\mathrm{i}qx}$ are linearly independent, all contributing modes, i.e., all modes where $\tilde{a}(q)\neq 0$ satisfy $-\mu +3\sum_k |\tilde{a}(k)|^2-\frac{3}{2}|\tilde{a}(q)|^2+q^2=0$. We can therefore construct solutions as follows: We pick an arbitrary set of $N$ pairwise distinct contributing modes $q_1, ..., q_N$. Then the mean squares of their amplitudes $\tilde{a}_1,...,\tilde{a}_N$ are determined by the linear system
\begin{equation}\label{eq:gen_steady_state}
	\begin{pmatrix}
	\frac{1}{2}&1&\dots&1\\
	1&\frac{1}{2}&\dots&1\\
	\dots&\dots&\dots&\dots\\
	1&1&\dots&\frac{1}{2}
	\end{pmatrix}
	\begin{pmatrix}
	|\tilde{a}_1|^2\\|\tilde{a}_2|^2\\ \dots\\|\tilde{a}_N|^2
	\end{pmatrix}=
	\frac{1}{3}
	\begin{pmatrix}
	\mu-q_1^2\\
	\mu-q_2^2\\
	\dots \\
	\mu-q_N^2
	\end{pmatrix},
\end{equation}
and the phases are left arbitrary.  As the solutions have to fulfill $|\tilde{a}_i|^2>0$ for all $i$, for each configuration there exists a threshold value $\mu_t$, such that the solution exists if $\mu>\mu_t$. Note that the single-mode solution arises as the special case for $N=1$, i.e., $\frac{1}{2}|\tilde{a}_1|^2=\frac{1}{3}(\mu-q_1^2)$. To illustrate the procedure and its result, we also detail the case $N=2$ where Eq.~(\ref{eq:gen_steady_state}) reduces to
\begin{equation}
	\begin{pmatrix}
	\frac{1}{2}&1\\
	1&\frac{1}{2}
	\end{pmatrix}
	\begin{pmatrix}
	|\tilde{a}_1|^2\\|\tilde{a}_2|^2\end{pmatrix}=
	\frac{1}{3}
	\begin{pmatrix}
	\mu-q_1^2\\
	\mu-q_2^2
	\end{pmatrix},
\end{equation}
and is solved by
\begin{equation}
	\begin{pmatrix}\label{eq:amplitudes_N2}
		|\tilde{a}_1|^2\\
		|\tilde{a}_2|^2
	\end{pmatrix}=\frac{2}{9}\begin{pmatrix}
	\mu+q_1^2-2q_2^2\\
	\mu-2q_1^2+q_2^2
	\end{pmatrix}.
\end{equation}
For $q_1=-q_2=:q_s$ the two equations reduce to 
\begin{equation}
	|\tilde{a}_{1,2}|^2=\frac{2}{9}(\mu-q_s^2).
\end{equation}
Notably, the condition for the existence of the solution is $\mu-q_s^2>0$ and is therefore the same as for the single harmonic mode with the same wavelength. 
Specifically, if $\tilde{a}_{1}=\tilde{a}_2$ the solution reads
\begin{equation}
	a(x,t)=\tilde{a}_s \cos(kx)\quad \text{with} \quad	|\tilde{a}_s|^2=\frac{8}{9}(\mu-q_s^2). 
\end{equation}
For this solution the original fields become
\begin{equation}
	\vecg u(x,t)=\sqrt{2}|\tilde{a}_s|\cos(kx)\begin{pmatrix}
	\sin (\omega q_s^2 t+\phi)\\\cos (\omega q_s^2 t+\phi)
	\end{pmatrix},
\end{equation}
which is a standing wave, where independently of space and time, the two fields are phase-shifted in time by $\frac{\pi}{2}$. Note that this state can be seen as a superposition of the left and right traveling waves, i.e., the  already discussed single-mode states. For the case $q_2> q_1$, Eq.~(\ref{eq:amplitudes_N2}) has a solution, when $\mu-2q_2^2+q_1^2>0$, i.e., it only exists for $\mu$-values that are strictly larger than the critical ones for the individual modes with wavenumbers $q_1$ and $q_2$. Further we note that any (multi-mode) steady state of the amplitude equation corresponds to a superposition of traveling waves in the full model. Here we call them oscillatory states. Fig.~\ref{fig:correspondence} gives a schematic overview of the discussed  states on the two levels of description, i.e., the full dynamic equation and the amplitude equation with real coefficient.

To analyze the stability of the single-mode states described above, we introduce a small complex perturbation $\epsilon a_p(x,t)$ with $\epsilon\ll 1$
\begin{equation}\label{eq:ansatz_perturb_ampl_eq}
\begin{split}
	a(x,t)&=\sqrt{\frac{2}{3}(\mu-q^2)}e^{\mathrm{i}qx}+\varepsilon a_p(x,t)\\
	\quad \text{with}\quad a_p(x,t)&\sim e^{\mathrm{i}kx+\lambda t}. 
	\end{split}
\end{equation}

\begin{figure}[htb!]
\centering
\includegraphics[width=0.7\hsize]{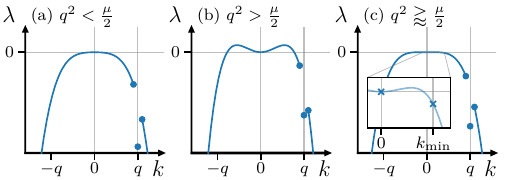}
\caption{Shown are selected dispersion relations of the single mode-solution according to Eq.~(\ref{eq:disp_single_mode}). Panels (a) and (b) show the stable and unstable case respectively. Panel (c) illustrates the effect of the discretized wavenumbers on the linear stability, i.e., in the displayed dispersion relation the wavenumber $q$ is unstable in the infinite-system limit but stable for a finite domain of length $2\pi/k_{\mathrm{min}}$. The term $\sim\delta_{k,q}$ in Eq.~(\ref{eq:disp_single_mode}) $q=k$ leads to a discontinuity at $k=q$, that is indicated by blue dots. }
\label{fig:dispersions_single_mode}
\end{figure}

Inserting into \eqref{eq:AE_og_scales} and linearizing in $\epsilon$ yields the dispersion relation
\begin{equation}\label{eq:disp_single_mode}
	\lambda(k)=k^2\left[-\delta_{k,q}(\mu-q^2)-\mu+2 q^2-k^2\right].
\end{equation}
The discontinuity, i.e., the term $\sim\delta_{k,q}$ arises from the evaluation of integrals ${\sim\int_{0}^{l} e^{\mathrm{i}(k-q)x}\mathrm{d} x}$ that correspond to Fourier transforms of harmonics and are related to the second nonlinear term in Eq.~\eqref{eq:AE_og_scales}, i.e., to the self-interaction of a traveling wave discussed in section~\ref{sec:ae}. Examples of dispersion relations are given in Fig.~\ref{fig:dispersions_single_mode}.

For a specific $q$ the single-mode solution is stable if the square bracket in Eq.~(\ref{eq:disp_single_mode}) is negative for all $k$.  
Since $\mu-q^2>0$ is necessary for the existence of the single-mode solution with wavelength $q$, the discontinuity is always stabilizing. Therefore, Eq.~\eqref{eq:disp_single_mode} has a large-scale instability if $-\mu+2 q^2-k^2>0$ for some $k$. In a finite system of size $l$ the spectrum of wavenumbers is discrete, such that the first allowed perturbation occurs at $k=k_{\text{min}}=2\pi/l$, see Fig.~\ref{fig:dispersions_single_mode} (c). In other words, $q$ is linearly stable if
\begin{equation}
  \label{eq:single_mode_stab}
	\quad q^2\leq\frac 12\left(\mu+k_{\text{min}}^2\right).
\end{equation}
In the limit of an infinite system, i.e., for $k_\text{min}\rightarrow 0$, the condition for linear stability reduces to
\begin{equation}
\quad q^2\leq \frac \mu 2.
\end{equation}
This implies that for any sufficiently large domain there exist several states with wavenumbers in the stable band, i.e., with $q\in [-\sqrt{\mu/2},\sqrt{\mu/2}]$ , i.e., we find a multistability of traveling waves of different wavelength in the full system. This is in stark contrast to the amplitude equation for the stationary large-scale instability with conservation law, i.e., the Cahn-Hilliard equation, where the only stable state has a structure length of system size.

The multistability of traveling wave solutions naturally raises the question of wavelength selection, i.e., as there exist stable single-mode states of various different wavenumbers, it is an intriguing question, which state a system is likely to realise as a function of the initial condition. To investigate this question and further validate our findings we continue with a numerical comparison of the dynamics as captured by the amplitude equation \eqref{eq:AE_og_scales} and by the full model \eqref{eq:minimal_CH}.

\section{Bifurcation analysis and time simulations}\label{sec:ae_sol_nuemeric}

For the numerical analysis we employ bifurcation techniques \cite{DWCD2014ccp,EGUW2019springer} and direct time simulations. Both are applied to the full dynamics \eqref{eq:minimal_CH}  and the amplitude equation \eqref{eq:AE_og_scales}. 

Remarkably, all steady states of the amplitude equation can be analytically determined as superpositions of (arbitrarily many) harmonics, as discussed in section~\ref{sec:ae_sol_analytic}.
As the NRCH dynamics contains various oscillatory states that are not easily tracked by numerical continuation, we restrict ourselves to simple steady states and traveling waves. As solution measures we employ the $L^2$-norms of the amplitude and of the original fields, i.e.,
\begin{equation} \label{eq:norm}
\begin{split}
||a||_{L^2}&=\left(\frac 2 l \int\limits_{-l/2}^{l/2} |a|^2\mathrm{d} x\right)^{\frac 1 2}\\
\text{and} \quad ||\vecg u||_{L^2}&=\left( \frac{1}{l}\int\limits_{-l/2}^{l/2} (u^2+v^2)\mathrm{d} x \right)^{\frac 1 2},
\end{split}
\end{equation}
respectively. To account for the relation \eqref{eq:ae_relation_og_scales} between $a$ and $\vecg {u}=(u,v)$, the norm of $a$ is scaled by a factor $\sqrt{2}$, such that the single-mode solution of the amplitude equation and the corresponding traveling wave in the full dynamics have the same norm. The numerical continuation of the NRCH dynamics is performed employing the finite-element library \textit{pde2path}.\cite{UeWR2014nmma}

For the time simulations, we use a pseudo-spectral, semi-implicit Euler method for both the amplitude equation and the full dynamics, where the time-step is adapted via a half-step method. To allow for a comparison we match the initial condition by projecting $\vecg u(x,t=0)$ onto the initial amplitude $a(x,t=0)$, apply the respective nonlinear time-evolution on the two levels of description, and then calculate the approximated solution from the amplitude equation $u_a(x,t)$ using the linear-Schrödinger relation \eqref{eq:ae_relation_og_scales}. 

\begin{figure}[htb!]
\centering
	\includegraphics[width=0.59\hsize]{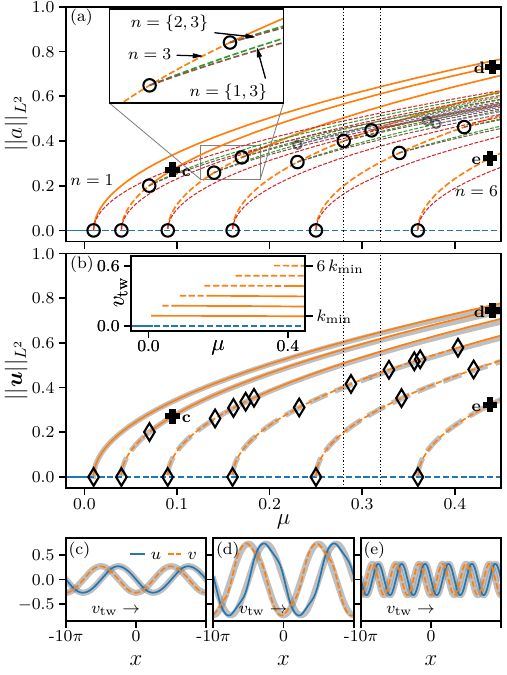}
	\caption{Bifurcation diagrams obtained for (a) the amplitude equation~\eqref{eq:AE_og_scales} and (b) the symmetric NRCH model~(\ref{eq:minimal_CH}). Solid [dashed] lines represent stable [unstable] states. The light gray lines in (b) reproduce the single-mode branches from (a) to allow for direct comparison.  The inset in panel (b) shows the velocity $v_\text{tw}$ of traveling waves. Large circles and diamonds mark pitchfork and Hopf bifurcations, respectively, while the small circles indicate bifurcations on tertiary (two-mode) branches. Panel (a) additionally includes cosine modes (red), the superposition of harmonic modes with two [three] distinct wavenumbers (green [purple]) and superpositions of cosine and harmonic modes (brown). 
Panels (c), (d) and (e) show $\vecg u$-profiles for the traveling wave states at loci marked by the crosses in (b), thereby the light gray curves reflect the real and imaginary part of the corresponding steady states in the amplitude equation (rescaled by a factor $\sqrt{2}$). The dotted vertical lines in (a) and (b) indicate the parameter values where the time simulations in Fig.~\ref{fig:spacetime_far_from_onset} are performed. The domain size is $l=20\pi$. }
	\label{fig:bif_analysis}
\end{figure}

We start with a discussion of the bifurcation structure. Fig.~\ref{fig:bif_analysis}(a) collects the analytical considerations from section~\ref{sec:ae_sol_analytic} in a bifurcation diagram whereas Fig.~\ref{fig:bif_analysis}(b) presents the homogeneous steady state and traveling wave states of the full model as obtained by path continuation. In Fig.~\ref{fig:bif_analysis}(a), at $\mu=0.01$ the homogeneous state becomes unstable with respect to the system-spanning mode with $k_\mathrm{min}=\frac{2\pi}{l}$ in a pitchfork bifurcation. 
Beyond the bifurcation, the homogeneous state has a fourfold degenerate positive eigenvalue,\footnote{Note that for a more transparent comparison with the real fields of the nonreciprocal Cahn-Hilliard equation we count real degrees of freedom. Therefore, every eigenvalue of the complex equation is twice degenerate as any complex perturbation corresponds to an arbitrary linear combination of real and imaginary part.} corresponding to real and imaginary perturbations with $\pm k_\mathrm{min}$. At the bifurcation, three branches emerge, corresponding to the linearly stable harmonic functions $a\sim e^{\pm\mathrm{i}k_\mathrm{min}x}$ and their unstable superposition $a\sim \cos(k_\mathrm{min}x)$. Note that the former two have identical norms as they are related by symmetry.

Similar bifurcations occur on the homogeneous steady state branch (primary branch) for all modes with $k_n= n k_\text{min}$, in Fig.~\ref{fig:bif_analysis} shown up to $n=6$, each further destabilising the trivial state (with fourfold degeneracy). Next, following the secondary $(n=2)$-branch, we observe another pitchfork bifurcation. At $\mu\approx0.06$ three tertiary branches emerge from the secondary branch, corresponding to the linear combinations of two modes with wavenumbers $n=1$ and $n=2$. There, the fourfold degenerate eigenvalue changes its sign from positive to negative and the secondary $(n=2)$-branch is stabilised. Something similar occurs for the secondary $(n=3)$-branch that emerges from the primary branch with two fourfold degenerate positive eigenvalues. Following the branch, we see that first the two-mode solutions with $n=\{1,3\}$ emerge stabilizing the first eigenvalue and second the two-mode solutions with $n=\{2,3\}$ emerge, stabilizing also this secondary branch, see inset in Fig.~\ref{fig:bif_analysis}(a). Similar successive stabilization occurs for every secondary branch, i.e., the $(n=i)$-branch is stabilized via $i-1$ consecutive pitchfork bifurcations.

Our analytic consideration allows us to find further branches -- in fact all branches -- for example, those with three contributing modes, that emerge from the ones with two contributing modes in further bifurcations, indicated by the small circles in Fig.~\ref{fig:bif_analysis}(a). 

\begin{figure*}[htb!]
	\includegraphics{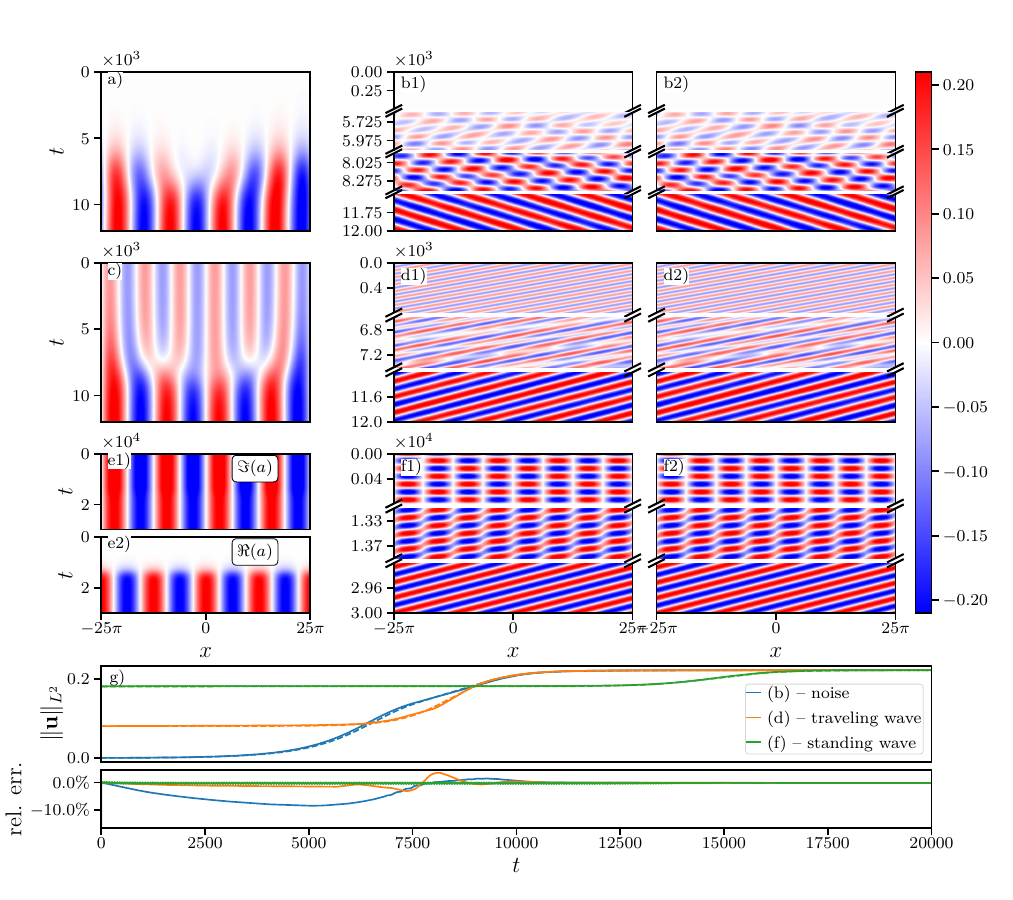}
	\caption{Spacetime plots for the amplitude (left column), the approximate full dynamics reconstructed from the amplitude equation (middle column) and the full dynamics obtained with the NRCH model (right column) with different initial condition, namely white noise (first row), an unstable traveling wave (second row) and a standing wave (third row). Panels (a), (c) and (e1) show $\Im(a)$, whereas (e2) shows $\Re(a)$. Panel (b), (d) and (f) display $u$. Panel (g)  shows the temporal evolution of the $L^2$-norm $\|\vecg u\|_{L^2}$ of the middle [right] column as dashed [solid] lines as well as their signed relative deviation. The parameters are $\mu=1/16$, $\omega=1$ and $l=50\pi$. The spatial discretisation consists of 512 points and the adapted timestep is usually around $1\times 10^{-2}$ for the full dynamics and $1\times 10^{-1}$ for the amplitude equation. }
	\label{fig:spacetime.pdf}
\end{figure*}

Next, turning to the bifurcation diagram of the corresponding NRCH system displayed in Fig.~\ref{fig:bif_analysis}(b), we see that each pitchfork bifurcation of the trivial branch in Fig.~\ref{fig:bif_analysis}(a) exactly corresponds to a Hopf bifurcation in Fig.~\ref{fig:bif_analysis}(b). At each bifurcation four complex eigenvalues cross the imaginary axis, corresponding to two identical pairs of complex conjugate eigenvalues, one for the left- and one for the right-traveling wave. Despite our technical restriction to the continuation of traveling wave states, it is obvious that another branch will emerge that corresponds to standing wave states formed by the superposition of right- and left-traveling wave. In fact the degeneracy is a general consequence of the translation- and parity invariance of the system, i.e., it is guaranteed by the underlying O(2)-symmetry and can be rigorously proven as the \emph{equivariant Hopf-theorem}. \cite{CrKn1991arfm, GoSt1985arma}

Following the emerging traveling wave branches, we see that, analogously to the amplitude equation, the $(n=1)$-branch is unconditionally stable, whereas branches for higher $n$ are at first unstable but stabilize in a sequence of Hopf bifurcations. With the exception of the $(n=2)$-branch, where the numerical precision is not sufficient to unambiguously tell, the degeneracy is partly lost, i.e., the $(n=3)$-branch stabilises in four consecutive Hopf bifurcations, where at each time only one complex conjugate pair of eigenvalues crosses the imaginary axis. In the case of the amplitude equation the degeneracy is due to the relative sign of the superposed harmonic mode having no impact, i.e., the superpositions of the contributing modes $\{k_2,k_1\}$ and $\{k_2,-k_1\}$ behave identically. In the case of the full dynamics this would mean that the superposition of a right-traveling wave with a left-traveling wave behaves analogously to the superposition of a right-traveling wave with a right-traveling wave. As the corresponding degeneracy is not seen in the full dynamics we speculate that it will also be lifted if the weakly nonlinear analysis is expanded up to a higher-order. 

Figs.~\ref{fig:bif_analysis} (c)-(e) show three examples of traveling wave states obtained by numerical continuation. The underlying grey lines show the steady states as obtained from the imaginary and real part of the corresponding solution of the amplitude equation, demonstrating that they also quantitatively match. Note that further away from the primary instability, the traveling waves are deformed, i.e., they do not have an exact harmonic shape as predicted by the amplitude equation. 
The inset of Fig.~\ref{fig:bif_analysis}(b) shows that the velocity of the traveling waves does not depend on the control parameter and scales linearly with the periodicity $n$. This quantitatively agrees with the dispersion relation that yields the traveling wave velocity $v_{\text{tw}}=\Omega(k_n)/k_{n}=\omega k_{n}$.

Next, we discuss time simulations performed close to the critical point, here at $\mu=1/16$, to verify that the amplitude equation also correctly reflects the transient dynamics from an arbitrary initial condition toward a stable traveling wave. We chose the domain size $l=50\pi$ in contrast to $l=20\pi$ in Fig.~\ref{fig:bif_analysis}, such that the previous analytical consideration predicts, that the $(n=4)$-traveling wave is the stable solution with the largest wavenumber. Fig.~\ref{fig:spacetime.pdf} (a)-(f) shows spacetime plots for various initial conditions, namely (top) white noise, (center) an unstable $(n=6)$-traveling wave solution and (bottom) an $(n=4)$-standing wave solution. The latter two are superposed with white noise of small amplitude ($\sim 10^{-3}$). Panels (b1), (d1) and (f1) show the field $u_{a}$ predicted from the amplitude equation, whereas panels (b2), (d2) and (f2) show the full dynamics of $u$. As we cannot visually resolve all oscillations over the whole simulation time, we show intermediate time intervals at critical times, i.e., when transitions occur. With the exception of (e) we only show $\Im(a)$ and $u$ because $\Re(a)$ and $v$ are qualitatively similar due to the underlying symmetries. For all initial conditions, the adapted timestep for the amplitude dynamics is persistently around one order of magnitude larger than for the full model, which roughly coincides with the ratio between fast and slow time scale of $\varepsilon^2=\mu=1/16$.

\begin{figure*}[htb]
	\includegraphics{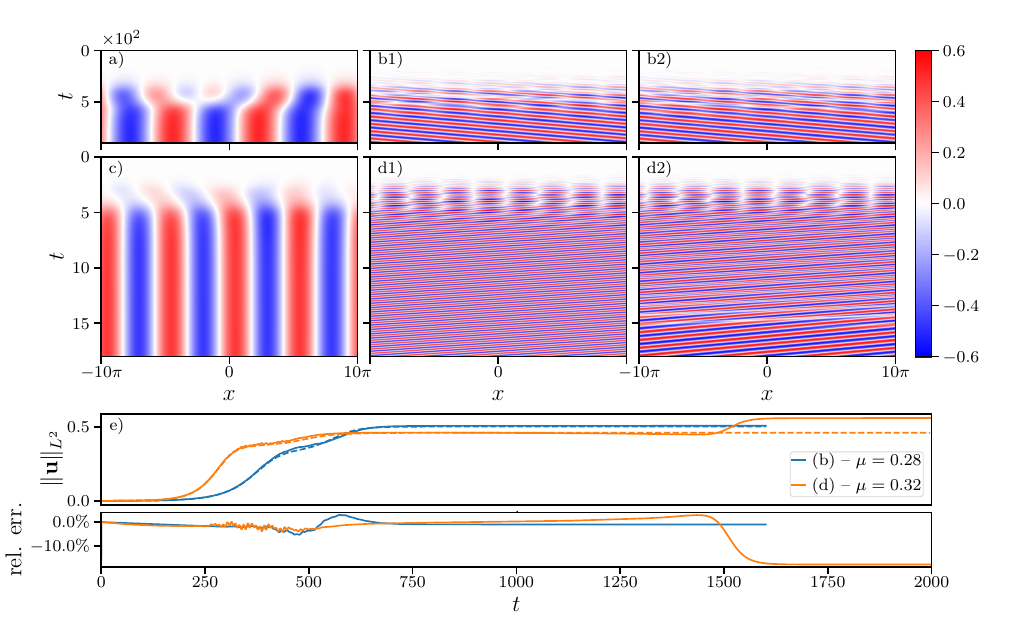}
	\caption{Spacetime plots for the amplitude $\Im(a)$ (panels (a) and (c)), the approximate dynamics for $u_{a}$ obtained from the amplitude equation (panels (b1) and (d1)) and the full dynamics $u$ of the NRCH model (panels (b2) and (d2)) with white noise initial condition ($\sim 10^{-3}$) at $\mu=0.28$ (first row) and $\mu=0.32$ (second row) and $l=20\pi$. Panel (e)  shows the temporal evolution of the $L^2$-norm $\|\vecg u\|_{L^2}$ of the middle [right] column as dashed [solid] lines as well as their signed relative deviation. The blue line that corresponds to panel (b) ends at $t=1600$ as there the numerical simulation has already converged to a stationary state. The spatial discretisation consists of 512 points and the adapted timestep is usually around $8\times 10^{-4}$ for the full dynamics and $3\times 10^{-3}$ for the amplitude equation. }
	\label{fig:spacetime_far_from_onset}
\end{figure*}

As $u$ and $u_{a}$ are visually indistinguishable in all panels, we conclude that the dynamics is well described by the amplitude equation, even after long simulation times. All the considered initial conditions settle to the ${(n=4)}$-traveling wave state. Simulations with white noise are repeated 10 times (not shown), and always yield the same final state, with left- and right-traveling waves appearing approximately equally often. This indicates that the basin of attraction of the $(n=4)$-solution is large compared to the one of the stable traveling wave states of lower periodicity. This is supported by the fact that it has the largest growth rate in the linear regime. 

In the case of the unstable $(n=6)$-traveling-wave in Fig~\ref{fig:spacetime.pdf} (c) and (d) one observes that it does not transition to an intermediate $(n=5)$-state, but after nearly staying the same until $t\approx 5 \times 10^3$ directly transitions into the $(n=4)$-state. The standing wave solution in panels (e) and (f) also remains nearly unchanged until at $t\approx 10^4$ it transitions into a traveling wave. As the standing wave initial condition here corresponds to a vanishing real part of the amplitude, real and imaginary part show qualitatively different behavior, with the imaginary part staying unchanged and the real part growing until it reaches the same magnitude at $t\approx 1.5\times 10^4$. As expected for a single-mode state, real and imaginary part of the amplitude are phase shifted by $\pi/2$. 

In Fig.~\ref{fig:spacetime.pdf}~(g), we compare the temporal evolution of the $L^2$-norm Eq.~\eqref{eq:norm} between the dynamics reconstructed from the amplitude equation and determined with the underlying NRCH model. The largest relative error $(\|u_a|\|_{L^2}-\|u|\|_{L^2})/\|u|\|_{L^2}$ is smaller than $10 \%$ and occurs during the transition regime at $t\approx 5000$, whereas it almost vanishes for the final states. 
To conclude the discussion of the behavior close to the onset of the instability, we emphasize that time simulations have to be treated with special care when judging the stability of time periodic states. Despite being unstable all intermediate states in panels (b), (d) and (f) of Fig.~\ref{fig:spacetime.pdf} stay qualitatively unchanged for many temporal periods, here up to 50 periods for the standing wave state. This is not unique to the conserved-Hopf bifurcation as any multiscale analysis aims to separate a slow transient at the onset of instability. However, here the linear Schrödinger-type relationship between fast and slow timescales causes intricate transient oscillations on the fast timescale. In contrast to this, in panels (a), (c), and (e) of Fig.~\ref{fig:spacetime.pdf}, i.e., on the slow timescale, the dynamics looks surprisingly simple.  

Finally, we briefly demonstrate that further away from the onset of the instability we start to see qualitative differences between the amplitude equation and the full dynamics. Following the previous discussion of the bifurcation analysis, we note in Fig.~\ref{fig:bif_analysis} that at $\mu=0.28$ both models predict the $(n=4)$-traveling wave state to be unstable. At $\mu=0.32$  though, the amplitude equation predicts it to have stabilised, whereas the full model predicts it to be still unstable (see the dotted vertical lines in Fig.~\ref{fig:bif_analysis}). The result of time simulations at these two $\mu$-values with white noise initial conditions is displayed in Fig.~\ref{fig:spacetime_far_from_onset}. The observed final states are in all cases in line with the stability results of the respective bifurcation analysis. 

Figs.~\ref{fig:spacetime_far_from_onset}(a,b) at $\mu=0.28$ retain the qualitative agreement between amplitude and full dynamics. Note that during the formation of the pattern, i.e., at $t\approx 350$ intermediate $(n=4)$-structures appear that due to their instability later give way to the final $(n=3)$-state. These $(n=4)$-structures similarly appear in Figs.~\ref{fig:spacetime_far_from_onset}(c,d) at $\mu=0.32$, with the difference that there they are stable and represent the final state obtained with the amplitude equation, but not with the full model.
Until $t\approx 500$ panels (d1) and (d2) look qualitatively similar. Although both then transition into an $(n=4)$-traveling wave state, there the full dynamics in panel (d2) starts to show a modulation, i.e., individual peaks become thinner [wider] and higher [lower] with a higher period than the traveling wave, until at $t\approx 1500$ two peaks merge and the dynamics settles to a stable $(n=3)$-traveling wave state. This is also visible in the temporal evolution of the $L^2$-norm in Fig.~\ref{fig:spacetime_far_from_onset}~(e), as for both cases the curves for reconstructed and full dynamics qualitatively agree. The relative error is below $5\%$ with the exception of the final state in (d), where the deviation is approximately $15\%$.

\section{Leading order amplitude equation: generic case}\label{sec:ae_generic}

As shown in Ref.~\onlinecite{FrTh2023prl}, a scalar (multi-field) model that features a conserved-Hopf instability, i.e., a large-scale oscillatory instability involving two conservation laws, can in the vicinity of the instability's onset always be reduced to a dynamical equation for two conserved scalar order parameter fields with nonlinearities up to third order in the fields. The NRCH model \eqref{eq:minimal_CH} we treated in section~\ref{sec:ae} has an additional \enquote{odd} field-exchange symmetry $(u,v)\to (-v,u)$. In consequence, the resulting nonlocal amplitude equation \eqref{eq:AE_og_scales} has a gradient dynamics structure that is not generic as will be shown next.

To treat the generic case we break the mentioned field-exchange symmetry by considering the system
\begin{equation}\label{eq:full_NRCH}
\begin{split}
	\partial_t u&=\partial_{xx}(-\sigma_1 u+u^3-\partial_{xx}u- (\rho+\alpha) v)\\
	\partial_t v&=\partial_{xx}(-\sigma_2 v+v^3-\kappa \partial_{xx}v-(\rho-\alpha) u),\\[.5em]
\text{with}&\quad \frac 1 l \int_0^l u \mathrm{d}x =\bar{u} \quad\text{and}\quad \frac{1}{l}\int_0^l v \mathrm{d}x =\bar{v}.
\end{split}
\end{equation}
Note that we do not use the general system obtained in Ref.~\onlinecite{FrTh2023prl}. Instead, to keep the calculations manageable, we only additionally introduce the rigidity ratio $\kappa$ and a reciprocal linear coupling of strength $\rho$ beside the nonreciprocal one of strength $\alpha$. Further, we use two different self-interactions $\sigma_1$ and $\sigma_2$. \footnote{Although, in principle, one could still eliminate one of the $\sigma$'s via a proper rescaling of time, we abstain from doing this as it is advantageous if both these parameters are able to smoothly cross zero. Specifically, all nondegenerate cases could be rescaled such that $\sigma_1=\pm1$.}

Furthermore, non-zero mean concentrations $\bar{u},\bar{v}$ break the field inversion symmetry $(u,v)\rightarrow (-u, -v)$ that equivalently can be broken by introducing quadratic nonlinearities, i.e., one may employ an affine transformation $(u,v)\rightarrow (u- \bar{u},v-\bar{v})$ to retain zero mean values but incorporate quadratic nonlinearities $3\bar{u}u^2$ and $3\bar{v}v^2$. 

Linearizing about the trivial state $(u,v)=(\bar{u},\bar{v})$ and expanding for $k\ll 1$ yields the dispersion relation \eqref{eq:dispersion} of the conserved Hopf instability, see appendix \ref{sec:dispersion}. In contrast to section~\ref{sec:ae}, here, the coefficients are
\begin{equation}\label{eq:parameters}
\begin{split}
  &\mu=\frac{\sigma_1+3\bar{u}^2+\sigma_2+3\bar{v}^2}{2},\qquad \delta'=\frac{1+\kappa}{2}\\
& \omega=\sqrt{\alpha^2-\rho^2-c^2} \quad \text{and} \quad \omega'=\frac{c(1-\kappa)}{2\omega}\\
 \text{where} \quad &c=\frac{\sigma_1+3\bar{u}^2-\sigma_2-3\bar{v}^2}{2}.\\
\end{split}
\end{equation}
Breaking the field inversion symmetry results in rather tedious calculations. Therefore, next we focus on the case of zero mean values, i.e., Eq.~\eqref{eq:full_NRCH} with $\bar{u}=\bar{v}=0$. The consequences of lifting this restriction are discussed in the conclusion.

\subsection{General case with field inversion symmetry}

Proceeding analogously to the minimal example in section~\ref{sec:ae}, we obtain an amplitude equation of similar form as the previous amplitude equation \eqref{eq:ae_relation_og_scales}. However, due to the more general setting, the inner gradient term and the nonlinearity pick up prefactors that algebraically depend on the parameters in Eq.~\eqref{eq:parameters}. Again, applying a Fredholm alternative at order $\varepsilon^5$, we obtain as amplitude equation 
\begin{equation}\label{AE:generic}
  \partial_t a=\partial_{xx}\left[-\mu a-\nu \partial_{xx}a+\frac{3}{2}\gamma \left(2a\langle |a|^2\rangle-\mathcal{F}^{-1}[\tilde a |\tilde a|^2]\right)\right]
\end{equation}
with the complex coefficients  
\begin{equation}\label{AE:generic:coeff}
	\nu=\delta'-\mathrm{i}\omega' \quad\text{and}\quad \gamma=1-\mathrm{i}\frac{c\rho}{\alpha\omega}
      \end{equation}
      where the expressions are derived in appendix~\ref{sec:field_inv_symmetry}. Note that the fields in Eq.~\eqref{eq:full_NRCH} are scaled such that the cubic nonlinearities have prefactor one. For the amplitude equation this implies $\Re{\gamma}=1$. 
      
The dynamics of the amplitudes $a(x,t)$ is related to the original dynamics via the linear Schrödinger-type relation
\begin{equation}\label{eq:ae_relation_og_scales_generic}
\quad \vecg{u}(x,t)=e^{\mathrm{i}\omega t\partial_{xx}}a(x,t)\vecg e_++e^{-\mathrm{i}\omega t\partial_{xx}}a^*(x,t)\vecg e_- \quad \text{with}\quad \vecg e_{\pm}=\sqrt{\frac{\alpha-\rho}{2\alpha}}\begin{pmatrix}\frac{c\pm i\omega}{\alpha-\rho}\\1
	\end{pmatrix}. 
\end{equation}
Here we only discuss key properties of equation~\eqref{AE:generic}, and leave an exhaustive study of its solutions and their stability for the future. First, we note that linearising~\eqref{AE:generic} about the trivial solution $a=0$, we obtain the dispersion relation $\lambda(k^2)=k^2\mu-\nu k^4=k^2\mu-\delta'k^4-\mathrm{i}\omega'k^4$, which corresponds to Eq.~\eqref{eq:full_NRCH} without the  fast timescale oscillation $\sim \mathrm{i}\omega k^2$. In other words, again, the fast timescale is fully separated and the equation only captures the slow timescale dynamics that may now include oscillations. 

In generic cases, $\nu$ and $\gamma$ are complex parameters and therefore the gradient dynamics structure shown by equation \eqref{eq:AE_og_scales} is lost in Eq.~\eqref{AE:generic}. However, Eqs.~\eqref{eq:parameters} and \eqref{AE:generic:coeff} reveal two nongeneric cases, where the amplitude equation has a gradient dynamics structure and reduces to Eq.~\eqref{eq:AE_og_scales}: First, if simultaneously reciprocal interactions are absent ($\rho=0$) and the two fields have equal rigidity ($\kappa=1$), and second if the self-interactions are symmetric ($\sigma_1=\sigma_2$), see Table~\ref{tab:degeneracies}. 

Due to the complex coefficients, harmonic solutions \eqref{eq:harmonic_soluitons} of Eq.~\eqref{eq:AE_og_scales} become traveling wave states $a(x,t)=a_0 e^{\mathrm{i}(kx-ft)}$ for Eq.~\eqref{AE:generic}. Inserting this ansatz and treating real and imaginary part separately, we obtain
\begin{align}
\Re\eqref{AE:generic}:\quad 0&=-\mu+k^2\delta'+\frac{3}{2}|a_0|^2\label{eq:real_part_cGE}\\
	\Im\eqref{AE:generic}:\quad f&=k^2\left(k^2\omega'-\frac{3}{2}|a_0|^2\frac{c\rho}{\alpha\omega}\right)\label{eq:imag_part_cGE}
\end{align}
Hence, for a traveling wave with wavenumber $k$, the amplitude $|a_0|$ is determined by Eq.~\eqref{eq:real_part_cGE} and then the frequency $f$ is determined by Eq.~\eqref{eq:imag_part_cGE} and features quartic and quadratic dependencies on wavenumber. Further, Eq.~\eqref{eq:real_part_cGE} ensures supercriticality, i.e., all bifurcations on the trivial branch are supercritical as $|a_0|^2 = 2(\mu-\mu_c)/3$ with $\mu_c=k^2\delta'$. 

In terms of the fields $u$ and $v$ of the full nonreciprocal Cahn-Hilliard model \eqref{eq:full_NRCH}, the relative modulation strength $||u||_{L^2}/||v||_{L^2}$, as well as the phase shift $\Delta  \phi$ between $u$ and $v$ are encoded in the eigenvector $\vecg e_+$ in Eq.~\eqref{eq:ae_relation_og_scales_generic}, namely
\begin{align}
	\frac{||u||_{L^2}}{||v||_{L^2}}&=\left|\frac{c\pm i\omega}{\alpha-\rho}\right|=\frac{\sqrt{\omega^2+c^2}}{|\alpha-\rho|}=\frac{\sqrt{\alpha^2-\rho^2}}{|\alpha-\rho|}=\sqrt{\left|\frac{\alpha+\rho}{\alpha-\rho}\right|} \label{eq:measure1}\\
	\tan({\Delta \phi})&=\frac{\Im{\frac{c\pm i\omega}{\alpha-\rho}}}{\Re{\frac{c\pm i\omega}{\alpha-\rho}}}=\frac{\omega}{c}. \label{eq:measure2}
\end{align}

Hence, while purely nonreciprocal coupling ($\rho = 0$) results in identical modulation strength of $u$ and $v$, any reciprocal coupling 
($\rho \neq 0$) gives rise to a difference in modulation strength of the two original fields. Equally, while identical (symmetric) self-interactions ($\sigma_1=\sigma_2 \rightarrow c= 0$) result in a  phase shift between  $u$ and $v$ of $\pi/2$ any asymmetry in the coupling ($c\neq 0$) implies a phase shift that differs from $\pi/2$. Note that the introduced measures \eqref{eq:measure1} and \eqref{eq:measure2} are well behaved as for a conserved-Hopf instability always $|\alpha|>|\rho|$. The nongeneric cases and the resulting consequences for the parameters on the amplitude level and the dynamics are summarized in Table~\ref{tab:degeneracies}.

\begin{figure}[htb!]
\centering
\includegraphics{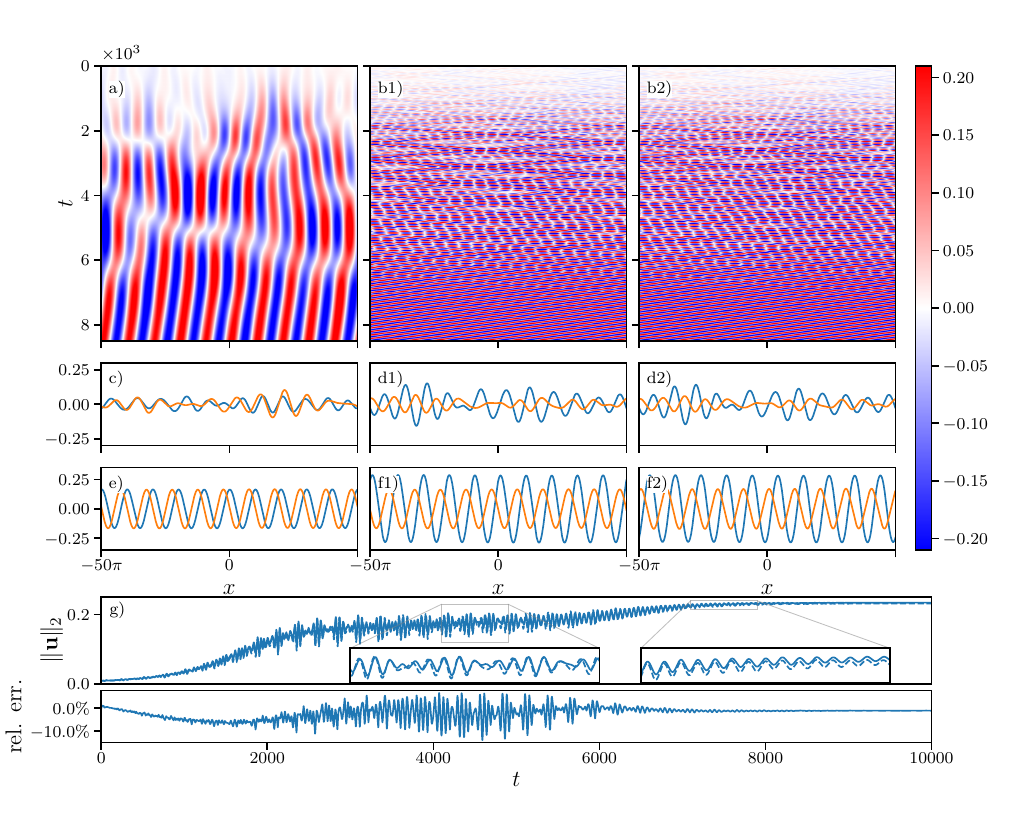}
\caption{Space-time plots of (a) the amplitude $\Im(a)$, (b1) the dynamics of $u_{1a}$ reconstructed from the amplitude equation and (b2) the full dynamics $u_1$ of the NRCH model, all with identical white noise initial condition ($\sim 10^{-1}$). Panels (c)-(f) show corresponding example profiles at (second row) $t=1.5\times 10^3$ and  (third row) $t=8\times 10^4$. The parameters are $\sigma_1=17/16$, $\sigma_2=-15/16$, $\rho=1$, $\alpha=2$ and $\kappa=0.1$. The corresponding parameters of the amplitude equation are $\mu=1/16$, $\nu=0.55+0.318\mathrm{i}$ and $\gamma=1+0.354\mathrm{i}$. Panel (g)  shows the temporal evolution of the $L^2$-norm $\|\vecg u\|_{L^2}$ of the middle [right] column as dashed [solid] lines as well as their signed relative deviation. The domain size is $l=100\pi$ with a spatial discretisation of 2048 points and the adapted timestep is usually around $1\times 10^{-2}$ for the full dynamics and $1\times 10^{-1}$ for the amplitude equation.}
\label{fig:spacetime_generic}
\end{figure}

To test our predictions for the generic case, we fix $\mu=1/16$ as in Fig.~\ref{fig:spacetime.pdf}, i.e., we stay close to the onset of the conserved-Hopf instability ($\varepsilon=1/4$), choose a generic set of parameters and a larger domain $l=100\pi$. The results are shown in Fig.~\ref{fig:spacetime_generic}. Again, the reconstructed dynamics obtained via Eqs.~\eqref{AE:generic} and \eqref{eq:ae_relation_og_scales_generic} shows excellent agreement with the full model \eqref{eq:full_NRCH}. Both, transients and the final states, coincide and only at very late times ($t\sim 8\times 10^4$, panels (f1) and (f2)) the reconstructed and the full dynamics have run slightly out of phase. Further, we find the relative modulation strength of the final state, $||u||_{L^2}/||v||_{L^2}=1.733$, to be in very good agreement with the theoretical prediction, here $\sqrt{\left|\frac{\alpha+\rho}{\alpha-\rho}\right|}=\sqrt{3}=1.732$. Similarly, a numerical estimate of the phase shift gives $\Delta \phi=2.160$,\footnote{Numerically, we calculate the phase shift as $\cos{\Delta\phi}=\frac{\frac{1}{l}\int_{-l/2}^{l/2} u v \mathrm{d}x}{||u||_{L^2}||v||_{L^2}}$, which becomes exact, if $u$ and $v$ are harmonic functions, i.e., $u=A\cos(kx+\Delta\phi)$, $v=B\cos(kx)$.} also in good agreement with the theoretical prediction of $2.186$ obtained from Eq.~\eqref{eq:measure2}. 

In contrast to the cases of Figs.~\ref{fig:spacetime.pdf} and \ref{fig:spacetime_far_from_onset}, the temporal evolution of the $L^2$-norms is nonmonotonic and oscillatory during the transient towards a traveling wave state [Fig.~\ref{fig:spacetime.pdf}~(g)] for both the reconstructed and the full dynamics. This is closely connected to the now absent gradient dynamics structure of the amplitude equation, i.e., due to the complex coefficients. This allows for oscillatory behavior also on the slow timescale. Although the relative error is sometimes up to $15\%$ and therefore larger than in the simple case treated in the previous sections, even the transient dynamics still qualitatively coincide (see insets in Fig.~\ref{fig:spacetime.pdf}~(g)). 

\begin{table}
\caption{ Summary of special (nongeneric) cases for the parameters of the NRCH model Eq.~\eqref{eq:full_NRCH} at $\bar{u}=\bar{v}=0$, the effect on the dispersion relation \eqref{eq:parameters}, and the resulting parameters in the amplitude equation \eqref{AE:generic:coeff}. This indicates that there are two independent limiting cases where \eqref{AE:generic:coeff} reduces to \eqref{eq:AE_og_scales}.} 
\label{tab:degeneracies}
\begin{tabular}{p{3.5cm} | p{2cm} p{2cm} p{2cm} p{3.1cm} c@{\,} p{2cm}}
\toprule
Description &Eq.\eqref{eq:full_NRCH}&Eq.\eqref{eq:parameters}&Eq.~\eqref{AE:generic:coeff}&Comment\\\hline\hline
Equal rigidity &$\kappa=1$&$\omega'=0$& $\nu\in \mathbb{R}$&&\ldelim\}{3}{*}&\multirow{2}{2cm}{\textbf{AE gradient \newline dynamics}} \\\cline{1-5}
Purely nonreciprocal coupling&$\rho=0$ &&$\gamma\in\mathbb{R}$&equal modulation \newline $||u||_{L^2}=||v||_{L^2}$\\\hline
Symmetric \newline self interaction &$\sigma_1=\sigma_2$&$c=0$ \newline $\omega'=0$&$\nu,\gamma\in \mathbb{R}$&\multicolumn{3}{p{5cm}}{\textbf{AE gradient dynamics} \newline Phase shift $\Delta\phi=\pm\pi/2$}\\\hline
\end{tabular}

\end{table}

\section{Conclusion}\label{sec:conclusion}
We have considered a large-scale oscillatory instability with conservation laws, i.e., the conserved-Hopf instability, that may occur in a wide spectrum of systems featuring (at least) two conservation laws. Examples of such systems include liquid two-layer films heated from below \cite{PBMT2005jcp} where spatio-temporal oscillation patterns may be readily observed, \cite{NeSi2017pf,NeSh2016jpat} and  a full cell polarity model \cite{OICK2007pcb} that represents a reaction-diffusion (RD) systems with more than one conservation law. Recently, RD systems with conservation laws gained renewed prominence as they are highly relevant for spatio-temporal protein patterns in biochemical processes. There, reactions result in the switching of protein conformations on a time scale where the overall protein density is conserved, e.g., MinE and MinD in ATP-driven Min oscillations. \cite{DKHH2018pnasusa,HaFr2018np,KYYF2019e,TYDF2022sa,BFLS2013pcb}
Further examples include two-species chemotactic systems, \cite{Wola2002ejam} mechanochemical waves in cytoskeleton and cytosol as modeled by active poroelastica, \cite{RAEB2013prl} oscillations in lipid-protein dynamics in cell membranes, \cite{JoBa2005pb} as well as liquid layers covered by self-propelled surfactants. \cite{PoTS2016epje}

Although such an oscillatory instability with conservation laws has been mentioned in the classification of instabilities in spatially extended systems by Cross and Hohenberg \cite{CrHo1993rmp} it has not yet been systematically studied. One may in general say that the role of conservation laws in pattern formation still holds quite a number of open questions. \cite{Knob2016ijam} A recently proposed alternative classification of instabilities in homogeneous isotropic systems into eight categories \cite{FrTh2023prl,FrTP2023ptrsapes} (based on three properties: large vs.\ small scale, stationary vs.\ oscillatory, and nonconserved vs.\ conserved) focused attention on the fact that weakly nonlinear theories (amplitude equations) were developed and analyzed for all of them with the exception of the conserved-Hopf instability. The general nonreciprocal Cahn-Hilliard model that was recently derived in Ref.~\onlinecite{FrTh2023prl} as a generic weakly nonlinear model valid in the vicinity of a conserved-Hopf instability (using $n$-field reaction-diffusion systems with two conservation laws as specific example to calculate all coefficients) is not an envelope equation, i.e., it does not reduce the considered time- and/or length scales by describing the time evolution of the amplitudes of temporal, spatial or spatio-temporal harmonic modes. Instead, as pointed out in the discussion of Ref.~\onlinecite{FrTh2023prl} it rather corresponds to an amplitude equation (of the zero-mode) on a higher level of a hierarchy of such universal equations valid in the vicinity of higher-codimension points. In consequence, it features other instabilities besides the conserved Hopf instability, namely, conserved Turing and Cahn-Hilliard instabilities (cf.\ conclusion and Supplementary Material of Ref.~\onlinecite{FrTh2023prl}). Amplitude equations that are not envelope equations can arise at specific points of higher codimension where no reduction scheme based on scale separation can be employed. This occurs, e.g., at transitions between an oscillatory and a stationary instability (there, the critical frequency approaches zero, i.e., no time-scale separation) or between a large-scale and a small-scale instability (critical wavenumber approaches zero, i.e., no length scale separation). The argument is further strengthened by the present work, since the complexity of the investigated NRCH models could only be reduced far away from a higher codimension point, in particular, far away from the transition point between Cahn-Hilliard instability and conserved-Hopf instability as there the NRCH model is its own amplitude equation. \cite{FrTh2023prl}  The general nonreciprocal Cahn-Hilliard model derived in Ref.~\onlinecite{FrTh2023prl} is closely related to the model proposed for oscillatory phase separation by F{\"o}rtsch and Zimmermann, as documented in Ref.~\onlinecite{Foertsch2023Bayreuth} (their GTOPS model in the forms of their Eqs.~(9.3) or (9.4), also called CHEOPS \cite{FoZi2023}).  The various models with simpler local contributions to the (non)equilibrium chemical potentials and constant mobilities analyzed in Refs.~\onlinecite{FrTP2023ptrsapes,FHKG2023pre,FrWT2021pre,FrTh2021ijam,SaAG2020prx,YoBM2020pnasusa,AlCB2023prl} all correspond to special cases of the model derived in Ref.~\onlinecite{FrTh2023prl} (see their Supplementary Material). The simplest version that still retains the conserved-Hopf instability, features a fully linearized second equation and purely linear couplings between fields \cite{Foertsch2023Bayreuth,SuKL2023pre,SuKL2023preb,BrMa2024prx}, and is called \enquote{Minimal Model} in \onlinecite{Foertsch2023Bayreuth}. It corresponds to the conserved equivalent of a FitzHugh-Nagumo reaction diffusion model.

In constrast to all these studies, the present work provides a derivation and first analysis of the amplitude equation for the conserved-Hopf instability on the lowest hierarchy level, i.e., the \enquote{missing} number eight. It is an amplitude equation that corresponds to an envelope equation, i.e., it properly reduces the considered time scales by describing the evolution on the slow time scale of the amplitude of fast harmonics. Now, the obtained versions of the equation can be employed to investigate in depth the universal behavior in the vicinity of the  instability onset. A key observation on the linear level has been that in the conserved case the slow and the fast dynamics of amplitude equation and original system, respectively, are related via a linear Schrödinger equation, i.e., reconstructing  the full multi-scale dynamics from the dynamics of the complex amplitude obtained in the weakly nonlinear model is similar to applying a quantum-mechanical time evolution operator in the case of a free-particle hamiltonian. It will be interesting  to investigate in the future whether this analogy can be further exploited.

The amplitude equation we have obtained as result of an intricate calculation up to fifth order. We have discussed it in two flavors: First, we have considered a nonreciprocal Cahn-Hilliard model with a field-exchange symmetry and purely nonreciprocal coupling and obtained the evolution equation~\eqref{eq:AE_og_scales} for the complex amplitude that features only real coefficients and has a gradient dynamics form.
Second, breaking the field-exchange symmetry, we have obtained the evolution equation~\eqref{AE:generic} for the complex amplitude that features complex coefficients and is nonvariational. In both cases we have still assumed that the two fields of the full model have zero mean value. In appendix~\ref{sec:app_quad_non}, we sketch the calculation for the case when this restriction is lifted. Conceptually, one again obtains an amplitude equation at order $\varepsilon^5$ with additional terms arising from the quadratic nonlinearity. However, as these terms are rather involved their practical use seems quite limited. The approach may now be used to analyze such equations for all mentioned versions of the nonreciprocal Cahn-Hilliard model and other more complicated models that show the conserved-Hopf instability.

Note that the operator-like relation between the levels of description was also noted by F{\"o}rtsch and Zimmermann. \cite{Foertsch2023Bayreuth,FoZi2023} However, in contrast to our exact approach resulting in equations~\eqref{eq:AE_og_scales} and \eqref{AE:generic}, they Taylor-expand and truncate the exponential operator to finally obtain a local amplitude equation with ten complex parameters [Eq.~(10.23) of Ref.~\onlinecite{Foertsch2023Bayreuth}] that is hard to analyze in general but still features analytic insights into certain simplified cases. It would be instructive to see a quantitative comparison of their reduced models and the dynamics of the full model.

The here-obtained amplitude equation has only a few parameters but features two types of nonlocal terms: one involves the mean value of a squared amplitude and the other is cubic in Fourier space, i.e., corresponds to a double convolution. Terms similar to the first one have also been discussed by Knobloch \cite{Knob1992} for counterpropagating wavetrains with velocities of order one in systems without conservation laws. There, the resulting amplitude equations are two complex Ginzburg-Landau equations with mean-field coupling. \cite{KnDe1990n} Such equations are also derived in the context of waves along certain combustion fronts.\cite{MaVo1992pd}

\begin{figure}[htb!]
\centering
	\includegraphics[width=0.65\hsize]{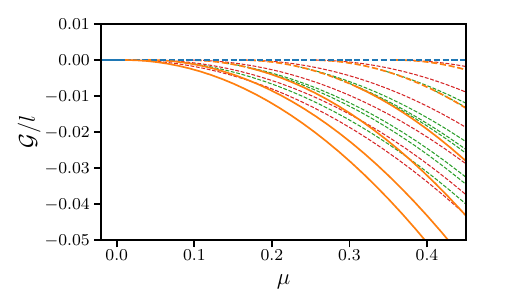}
	\caption{Energy density $\mathcal{G}/l$, i.e., Eq.~\eqref{eq:free_energy_ae}, for all steady states of the amplitude equation~\eqref{eq:AE_og_scales} that are included in Fig.~\ref{fig:bif_analysis}(a) shown here with identical line styles.}
	\label{fig:energy}
\end{figure}

Analyzing the two obtained amplitude equations we have found that the version with real coefficients has a variational structure, i.e., can be written as a gradient dynamics on the underlying energy functional~\eqref{eq:free_energy_ae}. Further, we have shown that all its steady states can be analytically obtained as combinations of harmonic modes with amplitudes given as solutions of an inhomogeneous linear system of equations. They correspond to various wave states of the full dynamics. The energy functional allows one to predict which wave state will ultimately be selected in a long-time limit. For wave number $q$ the single-mode states correspond to a mean energy density of $\mathcal{G}/l=-(\mu-q^2)^2/3$, i.e., the domain-filling mode always ultimately dominates. The full spectrum of all energies of steady states included in Fig.~\ref{fig:bif_analysis}(a) is given in Fig.~\ref{fig:energy} using identical line styles. We believe that two-mode state that bifurcates from a one-mode state at the respective final stabilizing pitchfork bifurcation corresponds to the threshold state that has to be overcome to transition to a state of lower energy. Therefore, the corresponding nondimensional energy difference $\Delta \mathcal{G}$ will allow one to obtain an estimate for the probability of remaining in a local energy minimum (as $1-\exp (-\Delta \mathcal{G})$).
Additionally, we have determined the linear stability of single-mode states. The obtained dispersion relation for perturbations of the single-mode state shows that it is a stationary large-scale instability not unlike the Eckhaus instability,\cite{Hoyle2006} however, with a different prefactor in the condition for onset and a different meaning of the wavenumber of the considered mode. For the Eckhaus instability it corresponds to the offset w.r.t.\ the finite critical wavenumber while here it is the offset w.r.t.\ zero.
A subsequent comparison of analytically and numerically obtained bifurcation diagrams for the steady states of the amplitude equation and the traveling waves of the full model, respectively, has has shown excellent agreement with only small differences emerging when moving away from the onset of the first mode. We have also compared transient dynamics for various initial conditions and also found very good agreement. Further, in our time simulations on the level of the amplitude equation we have observed the expected reduction in numerical effort as compared to the full model. The expected ratio of characteristic timesteps on the two levels of description roughly corrresponds to the factor $\varepsilon^2$ between fast and slow time scales. 

The generic amplitude equation with complex coefficients has a nonvariational structure. Here, we have only analytically determined single-mode traveling wave states of the amplitude equation. Further, we have compared the transient toward such a state as obtained from the reconstructed dynamics of the amplitude equation and from the full model. Also here the agreement has been very good. A discussion of further states and of their stability is left for the future. It will be particularly interesting to explore whether the derived complex equation allows for chaotic dynamics similar to its counterpart for nonconserved dynamics, the complex Ginzburg-Landau equation.\cite{ArKr2002rmp} 

In general, we believe that the approach can be further developed in several directions: On the one hand, employing a more general scaling (i.e., not fixing the prefactors of the cubic terms to one) one can assess how the amplitude equation simplifies when considering the above discussed simplest version of the NRCH model that still retains the conserved-Hopf instability, recently analyzed in some detail in Ref.~\onlinecite{BrMa2024prx}. One the other hand, further cubic coupling terms may be introduced into the derivation, e.g., to assess how the parameters of the complex amplitude equation change when considering the complex Cahn-Hilliard equation that is a limiting case of a model studied in Ref.~\onlinecite{Zimm1997pa} or other versions of the NRCH model considered in Refs.~\onlinecite{SaAG2020prx,YoBM2020pnasusa,FrWT2021pre,AlCB2023prl,SaGo2022arxiv}.

Finally, we point out that Ref.~\onlinecite{NeSh2016jpat} presents a review of mainly hydrodynamic models  with conservation laws that show large-scale oscillatory behavior \enquote{outside the world of the complex Ginzburg-Landau equation}. Specifically, similar oscillatory instabilities as the one discussed here occur in a liquid confined between two plates in Refs.~\onlinecite{Pism1988pra,NeSh2016jpat}, where the conservation of a concentration field and the approximate conservation of temperature in the limit of small Biot numbers can give rise to a large scale oscillatory instability. Refs.~\onlinecite{Pism1988pra,NeSh2016jpat} discuss several weakly and strongly nonlinear approaches, and it will be interesting to determine detailed relations to the here discussed real and complex versions of the amplitude equation for the conserved-Hopf instability. In particular, Ref.~\onlinecite{Pism1988pra} presents an amplitude equation that has a limiting case similar to the equation obtained and analyzed here.

Overall, the approach we have presented may prove useful to gain a more unified understanding of all mentioned cases of large-scale oscillatory pattern formation.

\section*{Acknowledgements}
We thank Tobias Frohoff-Hülsmann and Edgar Knobloch for valuable discussions. We would like to thank the Isaac Newton Institute for Mathematical Sciences, Cambridge, for support and hospitality during the program ``Anti-diffusive dynamics: from sub-cellular to astrophysical scales,'' where some work on this paper was undertaken. The program was supported by EPSRC grant EP/R014604/1. We also acknowledge fruitful discussions with many participants of the program and embedded workshops. Further, we acknowledge fruitful discussions in the context of the focus session \enquote{New Trends in Nonequilibrium Physics -- Conservation Laws
and Nonreciprocal Interactions} organized by Carsten Beta and Markus Bär at the \textit{DPG Spring Meeting 2024} in Berlin. DG thanks the \textit{Universitätsstiftung Münster} for financial support within the \textit{protalent} scholarship program.

\appendix
\section{Classification into eight instability types}\label{app:table}
\begin{table}[hbt]
\begin{tabular}{c || c | c}
& nonconserved dynamics& conserved dynamics\\
\hline
\hline
homogeneous/large-scale, stationary & Allen-Cahn (III$_\mathrm{s}$)& Cahn-Hilliard~(II$_\mathrm{s}$)\\
\hline
homogeneous/large-scale, oscillatory & Hopf\footnote{Also known as ``Poincar\'e-Andronov-Hopf''.} (III$_\mathrm{o}$)& conserved-Hopf~(II$_\mathrm{o}$)\\
\hline
small-scale, stationary & Turing (I$_\mathrm{s}$) & conserved-Turing (-)\\
\hline
small-scale, oscillatory & wave\footnote{Also called ``finite-wavelength Hopf'' or ``oscillatory Turing''.}  (I$_\mathrm{o}$)& conserved-wave (-)\\   
\hline
\end{tabular}
\caption{Overview of the classification and employed naming convention of linear instabilities and corresponding bifurcations of spatially extended homogeneous isotropic systems based on the three dichtomous properties: large-scale (L) vs.\ small-scale (S) instability, stationary (s) vs.\ oscillatory (o) instability, and nonconserved (N) vs.\ conserved (C) dynamics of relevant linear mode(s). For comparison, in parentheses the (incomplete) classification of Cross and Hohenberg \cite{CrHo1993rmp} is also given. Reprinted table with permission from Phys. Rev. Lett. 131, 107201 (2023)\cite{FrTh2023prl}. Copyright 2024 by the
American Physical Society. Also see the discussion in Ref.~\onlinecite{FrTP2023ptrsapes}.
}
\label{tab:linstab}
\end{table}

Table~\ref{tab:linstab} lists the eight instability types aluded to in Section~\ref{sec:intro} and their relation to the traditional classification. Each of eight types is characterized by the linear modes that dominate at and close to the corresponding instability threshold. Here, we also follow the naming convention given in the table. The traditional classification by Cross and Hohenberg \cite{CrHo1993rmp} only distinguishes six cases. It is included for reference. We remark that the statement ``type~II can often be scaled to resemble type~I'' on page~870 of Ref.~\onlinecite{CrHo1993rmp} is not correct, otherwise Cahn-Hilliard and Swift-Hohenberg equations would have similar dynamics close to onset. Furthermore, their remarks on frequencies in the context of the oscillatory case~II should be replaced by the present discussion in Section~\ref{sec:basics}.

\section{Calculations for the generic case}

In section~\ref{sec:ae} we have explained how to derive the amplitude equation \eqref{eq:AE_og_scales} as weakly nonlinear model in the vicinity of a conserved-Hopf bifurcation for a nonreciprocal Cahn-Hilliard model with a field-inversion symmetry and an additional field-exchange symmetry. Here, we generalize the procedure to the case where the field-exchange symmetry is broken, i.e., we show how to obtain the version with complex coefficients \eqref{AE:generic} as amplitude equation in the vicinity of the conserved-Hopf bifurcation for the more general nonreciprocal Cahn-Hilliard model \eqref{eq:full_NRCH}, however, with the condition of zero mean values of the two fields.

\subsection{Dispersion relation}\label{sec:dispersion}
Inserting the ansatz $\vecg u=\vecg u_0 + \varepsilon e^{\mathrm{i}kx+\lambda t}\delta \vecg u$, with $\vecg u_0=(\bar{u},\bar{v})$ into \eqref{eq:full_NRCH} and linearizing in $\varepsilon$ yields the linear eigenvalue problem
\begin{equation}
	\lambda	\delta\vecg u=k^2 \begin{pmatrix}
	\sigma_1+3\bar{u}^2 -k^2& (\rho+\alpha)\\
	(\rho-\alpha)&\sigma_2+3\bar{v}^2 -\kappa k^2
	\end{pmatrix}\delta \vecg u,
\end{equation}
with eigenvalues
\begin{equation}
\lambda_{1,2}(k^2)=k^2\,\left[\mu- \delta' k^2 \pm \sqrt{(c- \tilde c \,k^2)^2+\rho^2-\alpha^2}\right]
\end{equation}
where we introduced 
\begin{equation}
\begin{split}
&\mu=\frac{\sigma_1+3\bar{u}^2+\sigma_2+3\bar{v}^2}{2}\qquad \delta'=\frac{1+\kappa}{2}\\
&c=\frac{\sigma_1+3\bar{u}^2-\sigma_2-3\bar{v}^2}{2} \qquad \tilde c= \frac{1-\kappa}{2}
\end{split}
\end{equation}
The condition for large-scale oscillatory modes is a negative discriminant at $k^2\ll 1$, i.e., the contribution of the nonreciprocal coupling $\alpha^2$ has to be sufficiently large. Expanding the root for $k^2\ll 1$ we obtain
\begin{equation}
  \lambda_{1,2}(k^2)= k^2\,\left[\mu- \delta' k^2 \pm \mathrm{i}\left(\omega - \omega'\,k^2\right) \right]  
\end{equation}
where we furthermore introduced 
\begin{equation}
  \omega=\sqrt{\alpha^2-\rho^2-c^2}\qquad \omega'=\frac{c\tilde c}{\omega} \quad \text{and}\quad
\end{equation}
This corresponds to the dispersion relation~\eqref{eq:dispersion} without the $\mathcal{O}(k^6)$ terms that are not relevant for our calculation. The system is close to the onset of a conserved-Hopf instability if $|\mu|\ll 1$ and $\omega=\mathcal{O}(1)$.

\subsection{Case of intact field inversion symmetry}\label{sec:field_inv_symmetry}
First, we retain the field inversion symmetry by assuming $\bar{v}=\bar{u}=0$. Using the introduced parametrization, the system reads
\begin{equation}\label{eq:NRCH_abstract}
\begin{split}
\partial_t \vecg u &= \tens D \partial_{xx}\vecg u- \tens D_2
 \partial_{xxxx}\vecg u+\partial_{xx} \vecg N_3(\vecg u) - \mu\partial_{xx}\vecg u\\
 \text{with}\quad \tens D&=\begin{pmatrix}
	c &-(\rho+\alpha)\\
	-(\rho-\alpha) &-c
	\end{pmatrix}, \quad 
\tens D_2=\begin{pmatrix}
	1 & 0\\
	0 & \kappa
	\end{pmatrix},
 \quad \text{and} 
	\quad N_3(\vecg u)=\begin{pmatrix}u^3\\v^3\end{pmatrix}. 
\end{split}
\end{equation}
 Then,  we again formally expand the field in orders of $\varepsilon=\sqrt{|\mu|}$ and introduce the scalings $\partial_t=\varepsilon^2\partial_\tau+\varepsilon^4 \partial_T$ and $\partial_x=\varepsilon\partial_X$ with $\vecg u=\varepsilon \vecg u_1(X,\tau,T) + \text{h.o.t.}$. We start solving Eq.~(\ref{eq:NRCH_abstract}) order by order. The linear equation arising at $\mathcal{O}(\varepsilon^3)$ reads
\begin{equation}
	\tens L\vecg u_1=0 \quad \text{with} \quad \tens L=\tens 1 \partial_\tau -\tens D \partial_{XX}
\end{equation}
This equation is of linear Schrödinger-type and solved by
\begin{equation}\label{eq:amplitude_gen}
\begin{split}
\vecg{u}_1&=e^{\mathrm{i}\omega\tau\partial_{XX}}A(X,T)\vecg{e}_++c.c.\\
	\quad \vecg e_{\pm}&=\sqrt{\frac{\alpha-\rho}{2\alpha}}\begin{pmatrix}\frac{c\pm i\omega}{\alpha-\rho}\\1
	\end{pmatrix}, 
\end{split}
\end{equation}
where $\vecg e_\pm$ are normalized eigenvectors of $\tens D$ with eigenvalues $\pm \mathrm{i}\omega$. 
To apply the Fredholm alternative at higher orders we need the kernel of the adjoint linear operator $\tens L^\dagger=-\tens 1 \partial_\tau-\tensg D^\dagger\partial_{XX}$. The modes in the kernel are given by
\begin{equation}
\begin{split}
	\vecg m=e^{\mathrm{i}\omega\tau\partial_{XX}}B(X)\vecg e_+^\dagger+e^{-\mathrm{i}\omega\tau\partial_{XX}}C(X)\vecg e_-^\dagger \quad 
\text{with} \quad
	\vecg e_\pm^\dagger=\sqrt{\frac{\alpha+\rho}{2\alpha}}\begin{pmatrix}
	\frac{-c\pm\mathrm{i}\omega}{\alpha+\rho}\\1 
\end{pmatrix},
\end{split}
\end{equation}
where $\vecg e_\pm^\dagger$ are eigenvectors of $\tens D^\dagger$ with eigenvalues $\mp \mathrm{i}\omega$. Further note that $\langle\vecg e_+^\dagger ; \vecg e_- \rangle=0$. Hence, we can project Eq.~\eqref{eq:amplitude_gen} onto the amplitude $A(\hat{X},T)$ using $B(X)=\delta (X-\hat{X})$ and $C(X)=0$. 

At $\mathcal{O}(\varepsilon^5)$, we obtain the system
\begin{equation}\label{eq:order_eps4}
	\tens L \vecg u_3=-\tens D_2 \partial_{XXXX}\vecg u_1-\tilde{\mu}\partial_{XX}\vecg u_1-\partial_T \vecg u_1+\partial_{XX} \vecg N_3(\vecg u_1)
\end{equation}
As in the main text, we treat linear and nonlinear terms separately. For the linear terms, the Fredholm alternative yields 
\begin{equation}
\begin{split}
	0&=\langle e^{\mathrm{i}\omega\tau \partial_{XX}}\delta(X-\hat{X})\vecg e_+^\dagger; (-\tens D_2 \partial_{XXXX}\vecg u^{(1)}-\tilde{\mu}\partial_{XX}\vecg u^{(1)}-\partial_T \vecg u_1)\vecg e_+\rangle\\
\Leftrightarrow \quad \langle \vecg e_+^\dagger; \vecg e_+\rangle \partial_T A&=-\langle \vecg e_+^\dagger;\tens D_2 \vecg e_+\rangle\partial_{\hat{X}\hat{X}\hat{X}\hat{X}}A -\langle \vecg e_+^\dagger; \vecg e_+\rangle\tilde{\mu}\partial_{\hat{X}\hat{X}}A\,. 
\end{split}
\end{equation}
Next, we deal with the cubic contributions $\partial_{XX}\vecg N_3(\vecg u)$. Expressed by the amplitudes, they read
\begin{equation}\label{eq:ampl_nonl_u_app}
\begin{split}
\partial_{XX}\vecg N_3(\vecg u)=\partial_{XX}\begin{pmatrix}u_1^3\\v_1^3\end{pmatrix}
= \partial_{XX}\biggl[&\left(e^{\mathrm{i}\omega\tau\partial_{XX}} A(X,T)\right)^3\vecg w^{(1)} \\
+&3\left(e^{-\mathrm{i}\omega\tau\partial_{XX}} A^*(X,T)\right)\left(e^{\mathrm{i}\omega\tau\partial_{XX}} A(X,T)\right)^2 \vecg w^{(2)}\\
+&3\left(e^{-\mathrm{i}\omega\tau\partial_{XX}} A^*(X,T)\right)^2\left(e^{\mathrm{i}\omega\tau\partial_{XX}} A(X,T)\right)\vecg w^{(3)}\\
+&\left(e^{-\mathrm{i}\omega\tau\partial_{XX}} A^*(X,T)\right)^3\vecg w^{(4)}\bigg],
\end{split}
\end{equation}
where $\vecg w^{(i)}$ are given by
\begin{equation}
\begin{split}
\vecg w^{(1)}&=\begin{pmatrix}
(\vecg e_+)_1^3 \\ 
(\vecg e_+)_2^3 
\end{pmatrix},\quad 
\vecg w^{(2)}=\begin{pmatrix}
(\vecg e_+)_1^2 (\vecg e_-)_1\\ 
(\vecg e_+)_2^2 (\vecg e_-)_2
\end{pmatrix} \quad
\vecg w^{(3)}=\begin{pmatrix}
(\vecg e_+)_1 (\vecg e_-)_1^2\\ 
(\vecg e_+)_2 (\vecg e_-)_2^2
\end{pmatrix} \quad \text{and}\quad 
\vecg w^{(4)}=
 \begin{pmatrix}
(\vecg e_-)_1^3 \\ 
(\vecg e_-)_2^3 
\end{pmatrix}. 
\end{split}
\end{equation}

The Fredholm alternative for the term $\sim \vecg w^{(2)}$ then yields the contribution.
\begin{equation}
\begin{split}\label{eq:nonlinearity_proj_app}
&3\left\langle e^{\mathrm{i}\omega\tau\partial_{XX}}\delta(X-\hat{X})\vecg e_+^\dagger;\partial_{XX}\left(e^{-\mathrm{i}\omega\tau\partial_{XX}} A^*(X,T)\right)\left(e^{\mathrm{i}\omega\tau\partial_{XX}} A(X,T)\right)^2 \vecg w^{(2)}\right\rangle\\
=&\frac{3}{2}\left\langle e^{\mathrm{i}\omega\tau\partial_{XX}}\sum_Q e^{\mathrm{i}Q(X-\hat{X})}\vecg e_+^\dagger;\partial_{XX}\left(e^{-\mathrm{i}\omega\tau\partial_{XX}} \sum_K \Tilde{A}^*(K,T)e^{-\mathrm{i}KX}\right)\left(e^{\mathrm{i}\omega\tau\partial_{XX}} \sum_K \Tilde{A}(K,T)e^{\mathrm{i}KX}\right)^2 \vecg w^{(2)}\right\rangle\\
=&\frac{3 \Omega_{\text{min}}}{L} \langle \vecg e_+^\dagger; \vecg w^{(2)} \rangle\sum_{Q,K,K',K''} e^{\mathrm{i}Q\hat{X}} \Tilde{A}^*(K,T)\Tilde{A}(K',T)\Tilde{A}(K'',T)\left(-Q^2\right)\times\\&\times\int_0^{1/\Omega_{\text{min}}} e^{\mathrm{i}\omega\tau (Q^2+K^2-K'^2-K''^2)} \mathrm{d} \tau \int_0^L e^{\mathrm{i}X(-Q-K+K'+K'')} \mathrm{d} X \\
	=&3 \langle \vecg e_+^\dagger; \vecg w^{(2)} \rangle\left[2\sum_Q e^{\mathrm{i}Q\hat{X}} (-Q^2)\tilde{A}(Q,T) \sum_K \tilde{A}^*(K,T) \tilde{A}(K,T)-\sum e^{\mathrm{i}QX'} (-Q^2)\tilde{A}(Q,T) \tilde{A}(Q,T) \tilde{A}^*(Q,T)\right]\\
	=&3 \langle \vecg e_+^\dagger; \vecg w^{(2)} \rangle \partial_{\hat{X}\hat{X}}\left(2  \mathcal{F}^{-1}[\tilde{A}(K,T)\sum_{K'}|\tilde A(K',T)|^2]-\mathcal{F}^{-1}[\tilde{A}(K,T)|\tilde A(K,T)|^2] \right).
\end{split}
\end{equation}
Collecting all terms and introducing the original scales, we obtain the amplitude equation
\begin{equation}
	\partial_t a=\partial_{xx}\left[-\mu a-\nu \partial_{xx}a+\frac{3}{2}\gamma \left(2a\langle |a|^2\rangle-\mathcal{F}^{-1}[\tilde a |\tilde a|^2]\right)\right]
\end{equation}
where the general complex coefficients are given as
\begin{equation}
\begin{split}
	\nu &= \frac{1}{\langle \vecg e_+^\dagger; \vecg e_+ \rangle}\langle \vecg e_+^\dagger;\tens D_2 \vecg e_+\rangle=\delta'-\mathrm{i}\omega'\\
	\gamma&=\frac{2}{\langle \vecg e_+^\dagger; \vecg e_+ \rangle} \langle \vecg e_+^\dagger; \vecg w^{(2)} \rangle = 1-\mathrm{i}\frac{c\rho}{\alpha \omega}
\end{split}
\end{equation}
The particular coefficients arise from the NRCH model \eqref{eq:full_NRCH}, i.e.,

\begin{equation}
\begin{split}
\langle \vecg e_+^\dagger; \vecg e_+ \rangle&=\sqrt{\frac{\alpha+\rho}{2\alpha}}\sqrt{\frac{\alpha-\rho}{2\alpha}}\left(\frac{-c-\mathrm{i}\omega}{\alpha+\rho}\frac{c+\mathrm{i}\omega}{\alpha-\rho}+1\right)\\
&=\frac{\sqrt{\alpha^2-\rho^2}}{2\alpha}\frac{\omega^2-c^2+\alpha^2-\rho^2-2\mathrm{i}\omega c}{\alpha^2-\rho^2}\\
&=\frac{\omega^2-\mathrm{i}\omega c}{\alpha\sqrt{\alpha^2-\rho^2}}\\
\langle\vecg e_+^\dagger;\tens D_2 \vecg e_+\rangle&=\sqrt{\frac{\alpha+\rho}{2\alpha}}\sqrt{\frac{\alpha-\rho}{2\alpha}}\left(\frac{-c-\mathrm{i}\omega}{\alpha+\rho}\frac{c+\mathrm{i}\omega}{\alpha-\rho}+\kappa\right)\\
&=\frac{\sqrt{\alpha^2-\rho^2}}{2\alpha}\frac{\omega^2-c^2+\kappa(\alpha^2-\rho^2)-2\mathrm{i}\omega c}{\alpha^2-\rho^2}\\
&=\frac{\omega^2\frac{(1+\kappa)}{2}-c^2\frac{(1-\kappa)}{2}-\mathrm{i}\omega c}{\alpha\sqrt{\alpha^2-\rho^2}}\\
\implies \quad  \nu &= \frac{\omega^2\frac{(1+\kappa)}{2}-c^2\frac{(1-\kappa)}{2}-\mathrm{i}\omega c}{\omega^2-\mathrm{i} \omega c}\\&
=\frac{1}{\omega^4+\omega^2c^2}\left[\omega^4\frac{(1+\kappa)}{2}+\omega^2c^2\frac{(1-\kappa)}{2}+\omega^2c^2+\mathrm{i}\omega c \left(\omega^2\frac{(1+\kappa)}{2}-c^2\frac{(1-\kappa)}{2}-\omega^2\right)\right]\\
&=\delta'-\mathrm{i}\omega'\\
\langle \vecg e_+^\dagger; \vecg w^{(2)} \rangle &=\sqrt{\frac{\alpha+\rho}{2\alpha}}\sqrt{\frac{\alpha-\rho}{2\alpha}}^3\left[\frac{-c-\mathrm{i}\omega}{\alpha+\rho}\left(\frac{c+\mathrm{i}\omega}{\alpha-\rho}\right)^2\frac{c-\mathrm{i}\omega}{\alpha-\rho}+1\right]\\
&=\frac{\sqrt{\alpha^2-\rho^2}(\alpha-\rho)}{4\alpha^2}\left[\frac{(\omega^2+c^2)(\omega^2-c^2-2\mathrm{i}\omega c)}{(\alpha^2-\rho^2)(\alpha-\rho)^2}+1\right]\\
&=\frac{\sqrt{\alpha^2-\rho^2}}{4\alpha^2(\alpha-\rho)}\left[\omega^2-c^2-2\mathrm{i}\omega c+(\alpha-\rho)^2\right]\\
&=\frac{\sqrt{\alpha^2-\rho^2}}{2\alpha^2(\alpha-\rho)}\left[\alpha^2-c^2-\rho\alpha-\mathrm{i}\omega c\right]\\
\implies\quad\gamma&=\frac{2(\alpha^2-\rho^2)}{2\alpha(\alpha-\rho)}\frac{\alpha^2-c^2-\rho\alpha-\mathrm{i}\omega c}{\omega^2-\mathrm{i}\omega c}\\
&=\frac{\alpha+\rho}{\alpha\omega}\frac{1}{\omega^2+c^2}\left[\omega\alpha^2-\omega c^2-\rho\alpha\omega+\omega c^2+\mathrm{i}\omega c(\alpha^2-c^2-\rho\alpha-\omega^2)\right]\\
&=1-\mathrm{i}\frac{c\rho}{\alpha \omega}
\end{split}
\end{equation}

\subsection{Case of broken field inversion symmetry}\label{sec:app_quad_non}
Finally, we break the field inversion symmetry by assuming $\bar{v}\neq0$ and $\bar{u}\neq0$. We use an affine transformation $(\hat{u}(x,t),\hat{v}(x,t))=(u(x,t)-\bar{u},v(x,t)-\bar{v})$, where the new fields (with hat) represent the deviations from the average of the old fields. From Eq.~\eqref{eq:full_NRCH} we obtain the system 
\begin{equation}\label{eq:NRCH_mean_deviations}
\begin{split}
	\partial_t \hat{u} &= \partial_{xx}\left[(\sigma_1+3\bar{u}^2)\hat{u} + 3\bar{u} \hat{u}^2+\hat{u}^3-\partial_{xx}\hat{u} - (\rho+\alpha)\hat{v}\right]\\
\partial_t \hat{v} &= \partial_{xx}\left[(\sigma_2+3\bar{v}^3)\hat{v}+3\bar{v}\hat{v}^2 +\hat{v}^3-\kappa\partial_{xx}\hat{v} - (\rho-\alpha)\hat{u}\right]\\
\text{with}&\quad \frac 1 l \int_0^l \hat{u} \mathrm{d}x =0 \quad\text{and}\quad \frac{1}{l}\int_0^l \hat{v} \mathrm{d}x =0.
\end{split}
\end{equation}
where we already dropped all constant terms from the brackets as they are eliminated by the outer derivatives. 
Dropping the hats, the system reads
\begin{equation}\label{eq:NRCH_mean_deviations}
\begin{split}
\partial_t \vecg u &= \tens D \partial_{xx}\vecg u- \tens D_2
 \partial_{xxxx}\vecg u+\partial_{xx}\vecg N_2(\vecg u) +\partial_{xx} \vecg N_3(\vecg u) - \mu\partial_{xx}\vecg u\\
 &\text{with} \quad N_2(\vecg u)=\begin{pmatrix}3\bar{u} u^2\\3\bar{v}v^2\end{pmatrix}.
\end{split}
\end{equation}
The remaining expressions ($\tens D, \tens D_2, N_2(\vecg u)$)  are as in Eq.~\eqref{eq:NRCH_abstract},
i.e., the quadratic nonlinearity $\partial_{xx}\vecg N_2(\vecg u)$ is the only difference to the previous case.
The linear problem at $\mathcal{O}(\varepsilon)$ is the same as before. 
However, now we have to refine our ansatz to $\vecg u=\varepsilon \vecg u_1 +\varepsilon^2 \vecg u_2$ to deal with the inhomogeneity that appears at $\mathcal{O}(\varepsilon^4)$, where the corresponding equation reads
\begin{equation}\label{eq:order_eps4}
	\tens L \vecg u_2=\partial_{XX} \vecg N_2 (\vecg u_1)
\end{equation}

The Fredholm alternative at $\mathcal{O}(\varepsilon^4)$ is automatically fulfilled. To demonstrate that, we  pick an arbitrary quadratic term in the amplitudes that arises after inserting the amplitudes $\vecg u_1$ into $\vecg N_2(\vecg u_1(X,\tau,\tau'))$ and multiplying out the binomial.
\begin{equation}
\begin{split}\label{eq:nonlinearity_squared}
&\left\langle \vecg m;\partial_{XX}\left(e^{-\mathrm{i}\hat{\omega} \tau} A^*(X,T)\right)\left(e^{\mathrm{i}\hat{\omega} \tau} A(X,T)\right) \tilde{\vecg w}\right\rangle\\
=&\left\langle e^{\mathrm{i}\omega\tau\partial_{XX}}\sum_Q e^{\mathrm{i}Q(X-X')}\vecg e_+^\dagger;\partial_{XX}\left(e^{-\mathrm{i}\omega \tau \partial_{XX}} \sum_K \Tilde{A}^*(K,T)e^{-\mathrm{i}KX}\right)\left(e^{\mathrm{i}\omega \tau \partial_{XX}} \sum_K \Tilde{A}(K,T)e^{\mathrm{i}KX}\right) \tilde{\vecg w}\right\rangle\\
\sim&\sum_{Q,K,K'} e^{\mathrm{i}QX'} \Tilde{A}^*(K,T)\Tilde{A}(K',T)\left(-(Q)^2\right)\int e^{\mathrm{i}\omega\tau (Q^2+K^2-K'^2)}\mathrm{d} \tau \int e^{\mathrm{i}X(Q+K-K')} \mathrm{d} X. 
\end{split}
\end{equation}
Hence nonzero contributions have to fulfill
\begin{equation}
	Q^2+K^2-K'^2=0\qquad \text{and}\qquad Q+K-K'=0,
\end{equation}
which can only hold if either $Q=0$ and $K=K'$ or $K=0$ and $Q=K'$. The two terms do not contribute as the first one vanishes due to the factor $Q^2$ in \eqref{eq:nonlinearity_squared}  and the second one as $\tilde{A}^*(0,T)=0$. A similar argument holds for all contributions in $\vecg N_2$. Hence, the solvability conditon is always fulfilled. In other words, there is no nontrivial elastic $1\rightarrow 2$ or $2\rightarrow 1$ scattering process of particles with equal mass in one dimension.

To solve the inhomogeneous linear PDE \eqref{eq:order_eps4} at $\mathcal{O}(\varepsilon^4)$, we insert the Fourier-transformed amplitudes into $\vecg N_2$ and express the vectorial parts in the basis $\{\vecg  e_+,\vecg e_-\}$
\begin{equation}\label{eq:quadratic_nonl}
\begin{split}
\vecg N_2=&\begin{pmatrix}3 \bar{u}u_1^2\\3 \bar{v}v_1^2\end{pmatrix}\\
		=&\sum_{K'K''}A(K')A(K'')e^{\mathrm{i}X(K'+K'')}e^{-\mathrm{i\omega\tau(K'^2+K''^2)}}\begin{pmatrix}
		3\bar{u}(\vecg e_+)_1^2\\
		3\bar{v}(\vecg e_+)_2^2
		\end{pmatrix}\\
&+		2\sum_{K'K''}\tilde{A}(K')\tilde{A}^*(K'')e^{\mathrm{i}X(K'-K'')}e^{-\mathrm{i\omega\tau(K'^2-K''^2)}}\begin{pmatrix}
		3\bar{u}(\vecg e_+)_1(\vecg e_-)_1\\
		3\bar{v}(\vecg e_+)_2(\vecg e_-)_2
		\end{pmatrix}\\
&+	\sum_{K'K''}\tilde{A}^*(K')\tilde{A}^*(K'')e^{\mathrm{i}X(-K'-K'')}e^{\mathrm{i\omega\tau(K'^2+K''^2)}}\begin{pmatrix}
		3\bar{u}(\vecg e_-)_1^2\\
		3\bar{v}(\vecg e_-)_2^2
		\end{pmatrix}\\
		=&\sum_{K'K''}\tilde{A}(K')\tilde{A}(K'')e^{\mathrm{i}X(K'+K'')}e^{-\mathrm{i\omega\tau(K'^2+K''^2)}}(\eta_1 \vecg e_+ + \eta_2\vecg e_-)\\
&+		2\sum_{K'K''}\tilde{A}(K')\tilde{A}^*(K'')e^{\mathrm{i}X(K'-K'')}e^{-\mathrm{i\omega\tau(K'^2-K''^2)}}(\eta_3 \vecg e_+ + \eta_3^* \vecg e_-)\\
&+	\sum_{K'K''}\tilde{A}^*(K')\tilde{A}^*(K'')e^{\mathrm{i}X(-K'-K'')}e^{\mathrm{i\omega\tau(K'^2+K''^2)}}(\eta_2^* \vecg e_+ + \eta_1^* \vecg e_-),
\end{split}
\end{equation}
where the coefficients are given by
\begin{equation}
\begin{split}
\eta_1&=\frac{3\alpha}{\mathrm{i}\omega}(\bar{u}(\vecg e_+)_1^2(\vecg e_-)_2-\bar{v}(\vecg e_+)_2^2(\vecg e_-)_1)\\
\eta_2&=-\frac{3\alpha}{\mathrm{i}\omega}(\bar{u}(\vecg e_+)_1^2(\vecg e_+)_2-\bar{v}(\vecg e_+)_2^2(\vecg e_+)_1)\\
\eta_3&=\frac{3\alpha}{\mathrm{i}\omega}(\bar{v}(\vecg e_+)_2(\vecg e_-)_2(\vecg e_-)_1-\bar{u}(\vecg e_+)_1(\vecg e_-)_1(\vecg e_-)_2)\\ 
\end{split}.
\end{equation}
The occurence of the complex conjugated factors arises from the fact that $\vecg N_2\in \mathbb{R}^2$.

We can then find a solution for each contribution, e.g., for the for the first term, we employ the ansatz
\begin{equation}
\vecg u_2 \sim \sum_{K'K''}f(K',K'')e^{\mathrm{i}X(K'+K'')}e^{-\mathrm{i\omega\tau(K'^2+K''^2)}}\vecg e_+
\end{equation}
Then, applying $\tens L$, we obtain
\begin{equation}
	\tens L \vecg u_2 = \sum_{K'K''}f(K',K'')i\omega \left[(K'+K'')^2-(K'^2+K''^2)\right] e^{\mathrm{i}X(K'+K'')}e^{-\mathrm{i\omega\tau(K'^2+K''^2)}}\vecg e_+
\end{equation}
and comparison to the first term in $\partial_{XX}\vecg N_2$ yields that 
\begin{equation}
	f(K',K'')=\frac{\eta_1}{\mathrm{i}\omega}\tilde{A}(K')\tilde{A}(K'')\frac{-(K'+K'')^2}{(K'+K'')^2-(K'^2+K''^2)}
\end{equation}
Therefore, we find the solution
\begin{equation}
\begin{split}
	\vecg u_2 =&\sum_{K',K''}\frac{1}{\mathrm{i}\omega}\bigg\{\tilde{A}(K')\tilde{A}(K'')e^{\mathrm{i}X(K'+K'')}e^{-\mathrm{i\omega\tau(K'^2+K''^2)}}\\
	&(K'+K'')^2\left[\frac{\eta_1}{(K'^2+K''^2)-(K'+K'')^2}\vecg e_++\frac{\eta_2}{(K'^2+K''^2)+(K'+K'')^2}\vecg e_-\right]\\
	&+2\tilde{A}(K')\tilde{A}^*(K'')e^{\mathrm{i}X(K'-K'')}e^{-\mathrm{i\omega\tau(K'^2-K''^2)}}\\
&(K'-K'')^2\left[\frac{\eta_3}{(K'^2-K''^2)-(K'-K'')^2}\vecg e_++\frac{\eta_3^*}{(K'^2-K''^2)+(K'-K'')^2}\vecg e_-\right]\\
&+\tilde{A}^*(K')\tilde{A}^*(K'')e^{\mathrm{i}X(-K'-K'')}e^{-\mathrm{i\omega\tau(-K'^2-K''^2)}}\\
&(K'+K'')^2\left[\frac{\eta_2^*}{(K'^2+K''^2)+(K'+K'')^2}\vecg e_+ +\frac{\eta_1^*}{-(K'^2+K''^2)+(K+K'')^2}\vecg e_-\right]\bigg\}\\
=&\sum_{K',K''}\frac{1}{2\mathrm{i}\omega}\bigg\{\tilde{A}(K')\tilde{A}(K'')e^{\mathrm{i}X(K'+K'')}e^{-\mathrm{i\omega\tau(K'^2+K''^2)}}\\
	&(K'+K'')^2\left[-\frac{\eta_1}{K'K''}\vecg e_++\frac{\eta_2}{K'^2+K'K''+K''^2}\vecg e_-\right]\\	&+2\tilde{A}(K')\tilde{A}^*(K'')e^{\mathrm{i}X(K'-K'')}e^{-\mathrm{i\omega\tau(K'^2-K''^2)}}\\
&(K'-K'')\left[\frac{\eta_3}{K''}\vecg e_+-\frac{\eta_3^*}{K'}\vecg e_-\right]\\
&+\tilde{A}^*(K')\tilde{A}^*(K'')e^{\mathrm{i}X(-K'-K'')}e^{-\mathrm{i\omega\tau(-K'^2-K''^2)}}\\
&(K'+K'')^2\left[\frac{\eta_2^*}{K'^2+K'K''+K''^2}\vecg e_+ +\frac{\eta_1^*}{K'K''}\vecg e_-\right]\bigg\}
\end{split}
\end{equation}

At $\mathcal{O}(\varepsilon^5)$ the relevant contributions arise from the combinations of the first term with $A^*$ and from the second term with $A$. These read
\begin{equation}
\begin{split}
2\begin{pmatrix}
3\bar{u}u_1 u_2\\
3\bar{v}v_1 v_2
\end{pmatrix}=&\frac{1}{2\mathrm{i}\omega}\sum_{KK'K''}\tilde{A}^*(K)\tilde{A}(K')\tilde{A}(K'')e^{\mathrm{i}X(-K+K'+K'')}e^{-\mathrm{i}\omega\tau(-K^2+K'^2+K''^2)}\\
&(K'+K'')^2\left[\frac{\eta_1}{K'K''}\vecg y_1
+\frac{\eta_2}{K'^2+K'K''+K''^2}\vecg y_2\right]\\
+&\frac{1}{\mathrm{i}\omega}\sum_{KK'K''}\tilde{A}(K)\tilde{A}(K')\tilde{A}^*(K'')e^{\mathrm{i}X(K+K'-K'')}e^{-\mathrm{i}\omega\tau(K^2+K'^2-K''^2)}\\
&(K'-K'')\left[\frac{\eta_3}{K''}\vecg y_2-
\frac{\eta_3^*}{K'}\vecg y_1\right]+\dots\\
=&\frac{1}{2\mathrm{i}\omega}\sum_{KK'K''}\tilde{A}^*(K)\tilde{A}(K')\tilde{A}(K'')e^{\mathrm{i}X(-K+K'+K'')}e^{-\mathrm{i}\omega\tau(-K^2+K'^2+K''^2)}\\
&\left\{\left[\eta_1\frac{(K'+K'')^2}{K'K''}+2\eta_3^*\frac{K'-K}{K'}\right]\vecg y_1+\left[\eta_2\frac{(K'+K'')^2}{K'^2+K'K''+K''^2}+2\eta_3\frac{K'-K}{K}\right]\vecg y_2 \right\},
\end{split}
\end{equation}
Finally, we apply the Fredholm alternative, where only the combinations with $Q=K'$ and $K=K''$ or $Q=K''$ and $K=K'$ survive, and we again have to account for the overcounted contribution. We therefore obtain the additional contribution $N_{\mathrm{qn}}$ from the quadratic nonlinearity
\begin{equation}
\begin{split}
N_{\mathrm{qn}}=\frac{1}{\mathrm{i}\omega}\sum_{Q}Q^2 e^{\mathrm{i}Q \hat{X}} \tilde{A}(Q)\sum_K \tilde{A}(K)\tilde{A}^*(K)\\
\left\{\langle \vecg e_+^\dagger;\vecg y_1\rangle \left[\eta_1\frac{(Q+K)^2}{QK}+\eta_3^*\frac{Q-K}{Q}\right]+\langle \vecg e_+^\dagger;\vecg y_2\rangle \left[\eta_2\frac{(Q+K)^2}{Q^2+QK+K^2}+\eta_3\frac{Q-K}{K}\right]\right\}\\
-\frac{1}{2\mathrm{i}\omega}(4\langle \vecg e_+^\dagger;\vecg y_1\rangle \eta_1+\frac{4}{3}\langle \vecg e_+^\dagger;\vecg y_2\rangle \eta_2)\sum_Q Q^2e^{\mathrm{i}Q\hat{X}} \tilde{A}(Q)|\tilde{A}(Q)|^2 
\end{split}
\end{equation}
in the amplitude equation. 
Conceptually, these terms represent the proper contribution from the quadratic nonlinearity at $\mathcal{O}(\varepsilon^5)$. However, this is of little practical use, since these terms are both hard to analyze and to compute, arguably even more than the original dynamics. Therefore, one might consider further approximations (with or without asymptotical rigor) to obtain a proper contribution from the quadratic nonlinearity.

%

\end{document}